\documentclass[amsmath,twocolumn,tighten, times]{aastex702}
\usepackage{booktabs}
\renewcommand{\figsetstart}{}
\renewcommand{\figsetnum}[1]{}
\renewcommand{\figsettitle}[1]{}

\renewcommand{\figsetstart}{}
\renewcommand{\figsetnum}[1]{}
\renewcommand{\figsettitle}[1]{}

\begin{document}

\title{Soft Excess Variability and a Test of X-ray Reprocessing in AGN Mrk\,359 and NGC\,4051 using NICER and TESS}

\author[0009-0002-9371-3036]{Tejas Sewak}
\affiliation{Department of Physics \& Astronomy, Texas Tech University,  Lubbock TX, 79410-1051, USA}
\email[show]{Tejas.Sewak@ttu.edu} 

\author[0000-0002-9113-7162]{Michael M. Fausnaugh}
\affiliation{Department of Physics \& Astronomy, Texas Tech University, Lubbock TX, 79410-1051, USA}
\email{Michael.Fausnaugh@ttu.edu} 

\author[0000-0002-8294-9281]{Edward M. Cackett}
\affiliation{Wayne State University, Department of Physics \& Astronomy, 666 W Hancock St, Detroit, MI 48201, USA}
\email{ecackett@wayne.edu}

\author[0000-0003-3828-2448]{Javier A. García}
\affiliation{NASA / Goddard Space Flight Center, Greenbelt, MD 20771, USA} \affiliation{California Institute of Technology, Pasadena, CA 91125, USA}
\email{javier@caltech.edu}

\author[0000-0003-1728-0304]{Keith Horne}
\affiliation{SUPA School of Physics and Astronomy, North Haugh, St\,Andrews, KY16 9SS, Scotland, UK}
\email{kdh1@st-andrews.ac.uk}

\author[0000-0002-7778-3117]{Rahul Jayaraman}
\affiliation{Department of Astronomy, Cornell University, Ithaca, NY 14850, USA}
\email{rj438@cornell.edu}

\author[0000-0002-0581-6506]{Viraja C. Khatu}
\affiliation{Canada-France-Hawai`i Telescope Corporation, Kamuela, HI 96743, USA}
\email{khatu@cfht.hawaii.edu}

\author[0000-0003-1183-1574]{Ethan Partington} 
\affiliation{Western Kentucky University, Bowling Green, KY 42101, USA}
\email{ethan.partington-aff@wku.edu}

\author[0000-0002-1173-7310]{Raj Prince}
\affiliation{Department of Physics, Institute of Science, Banaras Hindu University, Varanasi, 221005, India}
\email{priraj@bhu.ac.in}

\begin{abstract}

Continuum reverberation mapping of AGN accretion disks assumes that X-ray reprocessing drives the UV/optical variations on short timescales. This reprocessing should also produce an X-ray reflection spectrum, which could be the soft excess observed in many Type 1 AGN. Consequently, if the soft excess traces the reflected emission, the soft excess and optical light curves should be correlated because their variability is driven by the same illuminating X-ray continuum. We present results from a cross-correlation study using simultaneous NICER and TESS observations of $\sim$$30\mathrm{~days}$ for two AGN: Mrk\,359 and NGC\,4051. We performed time-resolved spectroscopic fits to the X-ray data using a model with power-law and soft-excess components. We find no strong correlation between the optical and soft excess light curves, with correlation coefficients $R(\tau)$\,$\sim$0.5 for both sources. The total X-ray light curve also shows a similarly poor correlation with the optical variations. However, we find a stronger correlation between the primary X-ray flux and the soft excess flux for both sources. The lack of correlation between the primary X-rays and soft-excess with the optical variations implies that simple linear X-ray reprocessing is not the dominant process driving the optical variability. Either an intervening reprocessor between the inner X-ray region and outer disc is present that dilutes the X-ray--optical correlation, or the optical variability is driven by other mechanisms such as extreme-UV reprocessing or intrinsic disk variability. 

\end{abstract}

\section{Introduction}

An accreting supermassive black hole is responsible for the continuum emission from Active Galactic Nuclei (AGN). Standard accretion disk theory predicts an optically thick and geometrically thin structure \citep{1973A&A....24..337S, 1973blho.conf..343N}. In this model, each disk annulus emits as a blackbody, with an effective temperature that decreases with radius as $T \propto r^{-3/4}$. For a supermassive black hole at low/moderate accretion rates, the model predicts that the disk temperature peaks at $\sim$$10^{5} \mathrm{~K}$, naturally explaining the origin of the UV/optical emission observed in AGN.

However, AGN also emit strongly in X-rays. Their spectra primarily exhibit a power-law shape with a high-energy rollover ($\sim$50--300$ \mathrm{~keV}$), together with a Fe K$\alpha$ emission line and Compton reflection hump above $\sim$ $20 \,\mathrm{keV}$. The X-rays are also highly variable at short-time scales, both in flux and the spectral shape 
\citep{1975ApJ...199L.139W,1981MNRAS.194..987M}. This emission is generally attributed to inverse Compton scattering of thermal disk photons through a region of hot relativistic electrons ($\sim$$10^{9}\mathrm{~K}$) known as a ``corona". The rapid X-ray variability is probably due to a change in the balance between Compton heating and cooling as the flux of seed photons entering the corona varies \citep{1997ApJ...476..620H}. Observational constraints also provide insights into the coronal geometry. Reverberation lags and microlensing studies find a compact origin for the corona, with a characteristic size of $\sim$10--20\,$R_{\mathrm{g}}$, where $R_{\mathrm{g}} = GM/c^{2}$ \citep[e.g.,][]{2009Natur.459..540F, 2010MNRAS.401.2419Z,2012MNRAS.422..129Z,2012ApJ...756...52M, 2013MNRAS.434.1129K, 2013ApJ...769...53M,2014MNRAS.438.2980C, 2014A&ARv..22...72U, 2014ApJ...789..125B, 2015ApJ...806..258M}. In addition, IXPE polarization measurements suggest that the corona is aligned with the disk midplane \citep{2023MNRAS.523.4468G,2024MmSAI..95a..27G}.

Apart from the power law emission in hard X-rays (at energies $>2\mathrm{~keV}$),
most AGN also exhibit a soft excess feature in their X-ray spectra, the exact physical origin of which is still an open question \citep{1985ApJ...297..633S, 2016A&A...588A..70B, 2024MNRAS.530.1603B}. If the power law is extrapolated to energies below $\,2\mathrm{~keV}$, an excess flux is observed in the soft X-ray band above the power law. The spectral shape of the soft excess is similar over a wide range of accretion rates and black hole masses, ruling out a purely thermal disk origin for the soft excess \citep{2004A&A...422...85P, 2004MNRAS.349L...7G, 2005A&A...432...15P, 2009A&A...495..421B}. 

The two main theories used to explain soft excess are: (1) Comptonization of disk photons in a warm corona with an electron temperature $kT$\,$\sim$0.1--1$\,\mathrm{keV}$ \citep{2012MNRAS.420.1848D,2020A&A...634A..85P, 2024A&A...690A.308P}; and (2) ionized, relativistically blurred-reflection of primary X-rays from the inner accretion disk \citep{2006MNRAS.365.1067C,2016MNRAS.462..751G}. It may also be that a combination of both relativistic reflection and a warm corona produces the soft excess \citep{2012ApJ...758...67L, 2021ApJ...913...13X, 2022MNRAS.515..353X, 2024MNRAS.530.1603B}. The warm corona model predicts the presence of an optically thick ($\tau$\,$\sim$10--40) region of electrons at a lower temperature than the primary corona, which up-scatter photons from the disk to soft X-rays. In the reflection model, the primary coronal emission irradiates the accretion disk, producing fluorescent lines below $\sim$\,$2$$\mathrm{\,keV}$. The emission lines are blurred by relativistic effects and reproduce the soft excess spectral component. \citealt{2016MNRAS.462..751G} further showed that if the density of electrons in the disk is allowed to extend up to $10^{20}\, \mathrm{cm^{-3}}$, free-free emission also contributes significantly to the soft-excess. This high-density disk reflection model of \texttt{RELXILL} has been used to fit the soft-excess feature in $\sim$\,$35$ AGN \citep{2019MNRAS.489.3436J,2022MNRAS.513.4361M,2024MNRAS.534..608M, 2025arXiv250115380M}. 

The X-ray reflection model makes a crucial prediction about the UV/optical AGN flux. When the disk is irradiated by X-rays from the primary corona, a fraction of the flux is absorbed by the accretion disk and reprocessed at UV/optical wavelengths. Many variability studies, therefore, assume that short timescale UV/optical variability is driven by variations in the primary X-rays. However the low correlation found between X-ray and UV/optical emission in recent reverberation campaigns does not provide evidence to support the critical assumption that X-rays drive variability, even when strong correlations have been found in UV/optical \citep{2015ApJ...806..129E,2016ApJ...821...56F,2018MNRAS.480.2881M,2019ApJ...870..123E,2020ApJ...896....1C,2020MNRAS.498.5399H,2021ApJ...922..151K,2022MNRAS.512L..33V,2023ApJ...947...62K,2023ApJ...958..195C,2025MNRAS.541..642P}.

A strong test of the X-ray reflection/reprocessing model is available by combining spectroscopic and timing information. Since a fraction of the primary X-ray flux is absorbed by the disk while the remainder of the flux is reflected, we should expect the reflection spectral component and the UV/optical emission to vary coherently. Thus, if the soft excess is dominated by reflected emission, we expect the soft excess variations to be correlated with the UV/optical emission.

Recently, \citealt{2024ApJ...977...77P,2025ApJ...986...81P} performed time-resolved spectral modeling using NICER X-ray observations together with Swift UV monitoring for two AGN: NGC\,7469 and Fairall\,9. For NGC\,7469, the strong correlation between variability of all X-ray spectral components and the Swift UVW2 band suggested that a standard X-ray reprocessing model can explain both the soft excess and spectral-timing properties. Fairall\,9 has a weaker correlation between the primary X-ray continuum and the UV emission, but a stronger correlation between the soft excess and the UV light curve. The authors suggested that this is due to an intermediate reprocessor between the hot corona and the accretion disk, possibly a warm corona that both reprocesses the X-ray emission above $>$2\,keV and produces the soft excess. 

In this paper, we extend the sample to include Mrk\,359 and NGC\,4051. Using time-resolved X-ray spectral modeling of NICER observations together with simultaneous longer wavelength coverage from TESS, we test the linear X-ray reprocessing/reflection model and study the variability of the soft excess. The paper is organized as follows. Section\,\ref{sec: observations and data reduction} summarizes the sample, simultaneous X-ray and optical observations, and data reduction. Section\,\ref{sec:Time resolved Spectroscopy} presents the time-resolved spectroscopic fits of the X-ray observations, and Section\,\ref{sec:cc_analysis} presents the cross-correlation results. Section\,\ref{Discussion} discusses theoretical implications of our results and compares them with variability results of other AGN. Section\,\ref{sec: conclusions} gives the concluding remarks.

\section{Observations and Data Reduction} \label{sec: observations and data reduction}

\subsection{Sample}

In this paper, we test the disk-reprocessing/X-ray reflection model in two AGN: Mrk\,359 and NGC\,4051. Both sources host relatively low-mass SMBHs ($\sim$$10^{6} M_{\odot}$), exhibit substantial X-ray and optical variability over the 30-day monitoring period, and historically show a strong soft-excess, making them well suited for studying the connection between the soft excess and optical variability.

Mrk\,359 is a type-1 AGN with a black hole mass of $0.5-5\times 10^{6} M_{\odot}$, estimated using the stellar velocity dispersion inferred from the width of the [O III] emission line \citep{2001A&A...377...52W}. \citealt{1993A&A...274..105W} first suggested that Mrk\,359 has a soft excess using the ROSAT survey data. \citealt{2001MNRAS.327L..37O} then analyzed \textit{XMM-Newton} observations and confirmed the presence of a soft excess by extrapolating the best fit 2--10 keV power law ($\Gamma=1.84 \pm 0.02$) to lower energies. They found that about $50 \%$ of the flux in the 0.2--2 keV band is due to the soft excess, which they modeled with a phenomenological blackbody. They did not find evidence of significant neutral intrinsic absorption or of a warm absorber in the soft X-ray band. \citealt{2020A&A...640A..99M} later analyzed joint \textit{XMM-Newton} and \textit{NuSTAR} observations, and compared the warm corona and relativistic reflection in high-density accretion disk models to explain the origin of soft excess. They found a preference for a two corona model (warm corona + primary corona) over the reflection scenario, and suggested that a warm ($kT \sim 0.3\mathrm{keV}$) and optically thick ($20 \leq \tau \leq 29$) spectral component explains the soft excess in Mrk\,359. 

NGC\,4051 is a nearby galaxy with a low mass black hole of $0.5-2\times 10^{6} M_{\odot}$, determined through reverberation mapping \citep{2017ApJ...840...97F}. X-ray spectral variability of NGC\,4051 has been studied extensively, with observations showing high- and low-flux states on weekly timescales \citep{2004MNRAS.350...10P, 2006MNRAS.368..903P}. During the low flux states, the power law spectrum becomes hard, with a photon index reaching $\Gamma\,$$\approx$$\,1$, far outside the range of 1.7--2.2 common in Type-1 AGN \citep{1998MNRAS.301L...1G, 2003ApJ...595..656U}. The flat power law component can be explained either by the presence of a strong reflection component at hard X-rays \citep{2003ApJ...595..656U, 2006MNRAS.368..903P}, or due to the presence of a warm absorber \citep{2004MNRAS.350...10P}. Meanwhile, the spectrum shows a strong soft excess, making NGC\,4051 an ideal source for modeling distinct spectral components. Regarding its timing properties, \citealt{2013MNRAS.429...75A} used \textit{XMM-Newton} observations to find a UV/X-ray correlation of $R\sim$0.5, with the UV lagging the X-rays by $\sim$3\,ks, and consistent results with cross-spectral analysis. Based on the correlation results, they suggested that $\sim$25 per cent of the UV variance is caused by thermal reprocessing of the incident X-ray emission, even though the X-ray luminosity variations, in principle, are energetic to drive variations in the UV band.

\subsection{Observations}

The two sources—Mrk\,359 and NGC\,4051—were observed in X-rays using the NICER X-ray Timing Instrument (XTI). NICER operates in the $0.22$–$15.0~\mathrm{keV}$ range and offers a high signal-to-noise photon-counting capability, particularly in the energy range where the soft excess dominates. 

Mrk\,359 was observed by NICER from 2022 September 29 to October 29 (MJD 59851–59881). A total of 31 spectral epochs, with an average cadence of 1 day, were obtained for Mrk\,359. The average exposure of the spectral epochs was $\sim$\,2\,ks at an average source count rate of $\sim$\,8\,cts/s. NGC\,4051 was observed from 2022 March 1 to March 26 (MJD 59639–59664). A total of 51 spectral epochs at an average cadence of half a day were obtained for NGC\,4051. The average exposure of the spectral epochs was $\sim$\,1.3\,ks at an average source count rate of $\sim$\,29\,cts/s.

The Transiting Exoplanet Survey Satellite (TESS) observes in the $600-1000 \mathrm{~nm}$. A single TESS sector covers $24^\circ \times 96^\circ$ of sky, with images taken at a cadence of  $10 \mathrm{~mins}$ over $27 \mathrm{~days}$. Mrk\,359 was observed in sector 57 (MJD 59853-59882) and NGC\,4051 in sector 49 (MJD 59637-59664). 

For NGC\,4051, we also use the public \textit{u}\,band light curve data of Las Cumbres Observatory (LCO) global robotic telescope network, produced in \citealt{2026MNRAS.548ag642M} and as published on Zenodo \citep{doi:10.5281/zenodo.17242993}.

\subsection{NICER Data Reduction}

To analyze the NICER data for both sources, we utilized HEASOFT version $6.34$ \citep{Blackburn_1995} and the latest CALDB file, \texttt{XTI20240206}. We followed the standard procedure outlined in the NICER data analysis thread.\footnote{\url{https://heasarc.gsfc.nasa.gov/docs/nicer/analysis_threads/}}
We ran the task \texttt{nicerl2} with default settings to perform calibration and screening steps, producing cleaned event files. For both sources, we then created Good Time Interval (GTI) files for each observation. We excluded intervals with non-X-ray background flares, where the light curves in different energy bands showed ``flares" correlated with overshoot count rate. Once all cleaned time intervals were obtained, spectral products for each GTI were generated by running the task \texttt{nicerl3-spect} with default settings. The \texttt{nicerl3-spect} task produces grouped spectra and generates background files. We used the variable \textsc{SCORPEON} model to generate the spectral background files, which we fit simultaneously with the source-spectrum models (see \S\ref{sec:Time resolved Spectroscopy}). The \textsc{SCORPEON} model has been demonstrated to be optimal for AGN analysis \citep{2024ApJ...977...77P}. We grouped the spectra using the optimal binning algorithm with a minimum of 1 count per bin \citep{kaastra16}. We also created a time-averaged spectrum for each source over the entire monitoring campaign using the task \texttt{niobsmerge}.

\subsection{TESS Photometry} \label{subsection:TESS}

We obtained the optical light curves by performing photometry on the full-frame images (FFIs) available from the Mikulski Archive for Space Telescopes (MAST). The photometry was done using the same method as \citealt{2021ApJ...908...51F,2023ApJ...956..108F}. In brief, this method employs difference imaging and point-spread function (PSF) photometry using the ISIS software package \citep{1999ascl.soft09003A,2000A&AS..144..363A} to produce a light curve in units of differential counts per second. The reference images for difference imaging were constructed from the $\sigma$-clipped average of 20 FFIs taken at times with the lowest backgrounds. We used ISIS to fit the TESS pixel response function (PRF) to each difference image and integrated the fitted PRF model to measure the differential flux. 

For cross-correlation analysis with X-ray light curves, the absolute flux level does not matter, so we used the differential light curves in our analysis. We binned the TESS light curves to $0.5$ and $0.28$ days intervals for Mrk\,359 and NGC\,4051, respectively, to sample the optical wavelengths at half the average cadence of the X-ray observations. For NGC\,4051 we also removed data before MJD 59642.5 and between MJD 59649.5 and 59656.5, as the Earth and/or Moon were above the sunshade at these times, introducing a complicated, time-variable signal that might bias the flux measurements.

\section{Time resolved X-ray spectroscopic fits} \label{sec:Time resolved Spectroscopy}

We fit the spectra from each NICER epoch using XSPEC version \texttt{12.14.1} \citep{Arna96}. For spectral modeling, we utilized NICER’s full energy range of 0.22–15.0\,keV, which is suggested while using the \textsc{SCORPEON} background model. We used the C-statistic \citep{1979ApJ...228..939C} to evaluate the model likelihood and determine the best-fit model parameters for individual epochs, since most of the spectral bins have few counts and are not in the Gaussian regime. For the time-averaged spectra, where the counts per bin are sufficiently high for the Gaussian approximation to be valid, we used the $\chi^{2}$ statistic. We checked the ratio plots and corresponding reduced $\chi^{2}$ values in the energy range 0.3--8.0\,keV to assess whether the best-fit model adequately describes our data. We estimated parameter uncertainties using the Markov Chain Monte Carlo (MCMC) method and we report the central 68\% credible regions throughout. We used \texttt{vern} cross-sections \citep{1996ApJ...465..487V} and \texttt{wilm} abundances \citep{2000ApJ...542..914W} when modeling absorption in \texttt{XSPEC}.

The goal of our spectral modeling routine is to obtain separate light curves for the soft-excess and primary X-ray components. The primary X-ray source spectrum follows a power-law shape, whereas the soft excess is a bump above the power law at energies below $\sim$\,2\,keV. We tested two different models to fit the soft excess: the high-density disk reflection model \texttt{RELXILLCP} \citep{2016MNRAS.462..751G}, and a \texttt{blackbody} model with a variable temperature $kT$. \texttt{RELXILLCP} is a physical model that depends on source parameters, and has been shown to produce a soft-excess feature by keeping the accretion disk density as a free model parameter \citep{2016MNRAS.462..751G, 2019MNRAS.489.3436J,2022MNRAS.513.4361M}. On the other hand, the \texttt{blackbody} is a phenomenological model employed solely to characterize the soft excess feature, as has been done historically \citep{1985MNRAS.217..105A,1999ApJS..125..317L}, and recently for NICER data of other AGN \citep{zoghbi2023measuringsoftexcessregion, 2024ApJ...977...77P,2025ApJ...986...81P}.

\subsection{Mrk\,359}

\begin{figure*}[t]
\centering
\includegraphics[width=\textwidth]{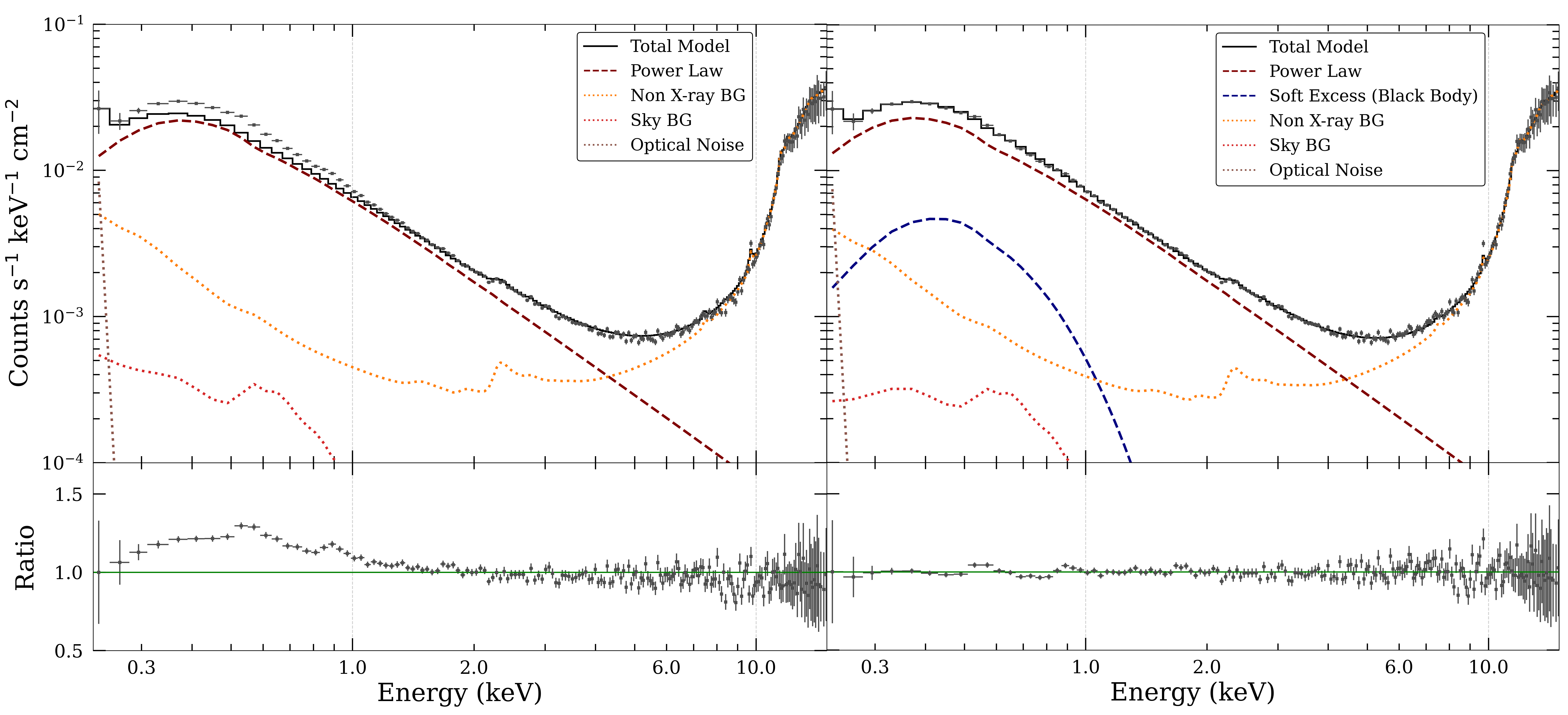}
\caption{Time-averaged X-ray spectrum of all NICER epochs for Mrk\,359. The gray points represent the observed spectrum (source + background), and the black solid line shows the total model spectrum (source model + background model). In the left panel, the maroon dashed line represents the best-fit \texttt{powerlaw} obtained in the energy range 2.0-15.0\,keV, modeled for the primary X-ray corona. The orange, red, and brown dotted lines represent the \texttt{SCORPEON} background model components. The \texttt{powerlaw} model fails to describe the data when extrapolated to energies below 2\,keV, as evident from the excess ratio of data to model in the lower panel plot. In the right panel, the best-fit combined model of \texttt{powerlaw} (maroon dashed line) and a \texttt{blackbody} component (blue dashed line) describes the data with $\chi^{2}_{\rm dof} = 134/114$ over 0.3--8.0\,keV energy range. The photon-index of the power law is $\Gamma = 1.98 \pm 0.01$ and $kT = 0.131\pm 0.002 \mathrm{~keV}$ for the \texttt{blackbody} component. The ratio of soft excess to primary X-ray integrated flux ($\Phi_{\mathrm{SE}}\textbf{/}\Phi_{\mathrm{PL}}$) in the energy range 0.3--8.0\,keV is $10^{-12.05}\textbf{/}10^{-10.90} \approx7\%$, which is smaller by a factor of $\sim$2 compared to that observed by \citealt{2020A&A...640A..99M}. 
\label{average_spectrum_mrk359}
}
\end{figure*}


\figsetstart
\figsetnum{2}
\figsettitle{Full set of NICER spectra and fitted models of Mrk\,359}

\figsetend

\begin{figure}[!ht]
  \centering
  \includegraphics[width=\columnwidth]{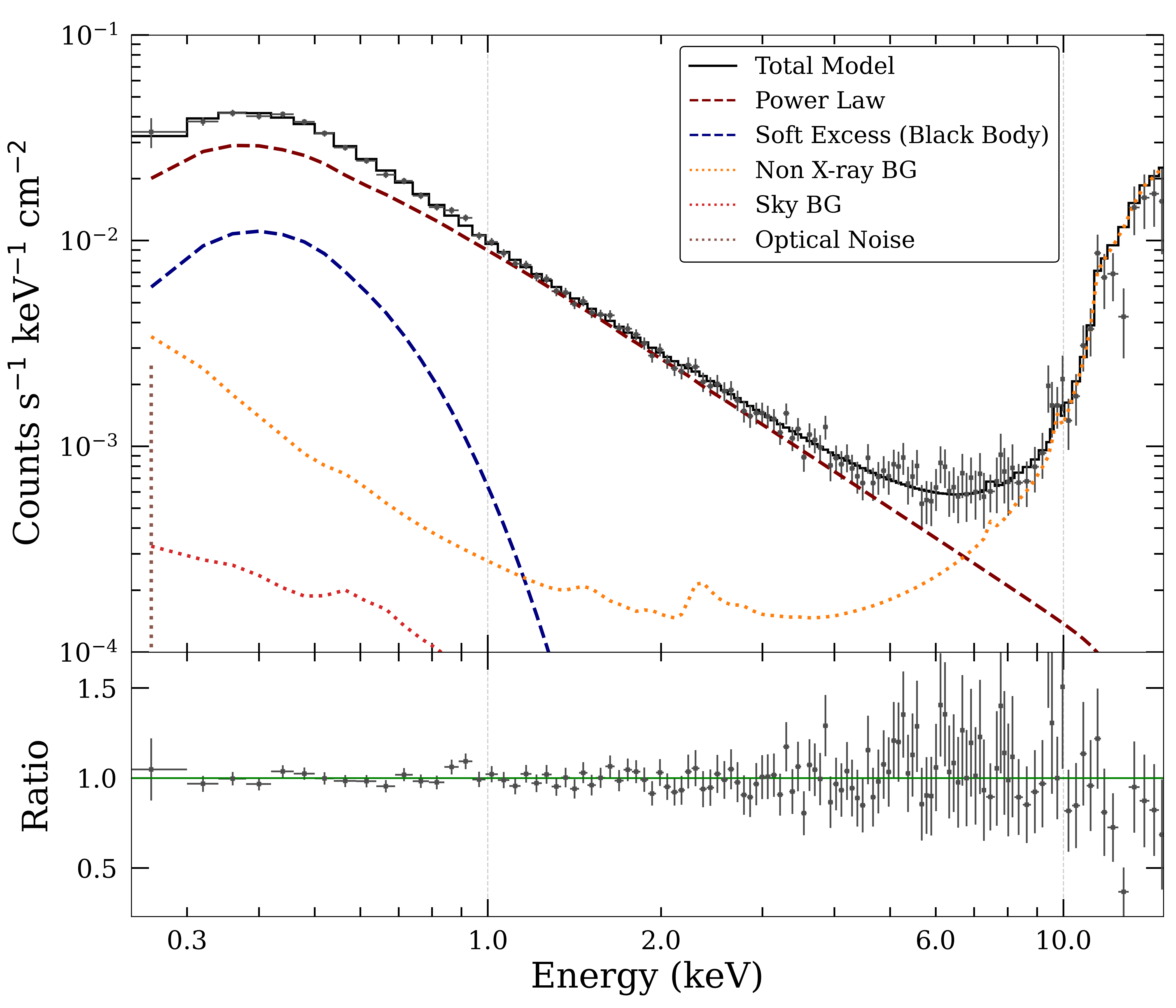} 
  \caption{X-ray spectrum from an individual NICER epoch of Mrk\,359, MJD\,59854.50, observed for a duration of 2.2\,ks. The ratio of soft excess to primary X-ray integrated flux ($\Phi_{\mathrm{SE}}\textbf{/}\Phi_{\mathrm{PL}}$) in the energy range 0.3--8.0\,keV is $10^{-11.77}\textbf{/}10^{-10.73} = 9.1\%$. The \texttt{powerlaw} photon-index is $\Gamma = 1.84^{+0.03}_{-0.02}$ and for the \texttt{blackbody} component $kT = 0.115\pm 0.004 \mathrm{~keV}$. The spectra and fitted models for all epochs are available as a figure set in the online journal.}

  \label{spectra_mrk359}
\end{figure}


To establish the presence of a soft-excess feature in Mrk\,359, we initially fit a single Galactic-absorbed power-law model over the 2--15 keV range. In XSPEC terminology, the spectral model is expressed as \texttt{tbabs\,$\times$\,powerlaw}. The hydrogen column density $N_{\rm H}$ for \texttt{tbabs} was fixed at  $4.38\times 10^{20}\mathrm{~cm}^{-2}$, as calculated using the HEASoft N$_{H}$ calculator and based on \citealt{2016A&A...594A.116H}. We found that the single absorbed \texttt{powerlaw} model provides a valid fit to the time-averaged spectrum $>2\rm \,keV$ with a $\chi^{2}/\mathrm{dof} = 77
/79$, and a photon index $\Gamma = 1.98^{+0.02}_{-0.06}$. Extrapolating the best-fit power law to lower energies, the time-averaged spectrum of Mrk\,359 clearly shows a soft excess, see Figure\,\ref{average_spectrum_mrk359}. 

We add a \texttt{blackbody} model to fit the shape of the soft excess, which gives a $\chi^{2}/\mathrm{dof} = 134/114$ in the energy range 0.3--8.0\,keV for the time-averaged spectrum. No significant positive residuals for the iron emission line are visible in Figure\,\ref{average_spectrum_mrk359}, and adding a narrow \texttt{zgauss} component improves the $\chi^{2}$ by only 4 for 2 additional $\rm dof$. The lack of a significant detection in our spectrum is more likely due to the limited S/N around the Fe\,K band.
We therefore do not include this additional component in our subsequent fits. In XSPEC terminology, our final spectral model is thus expressed as \texttt{tbabs\,$\times$\,(cflux\,$\times$\,powerlaw\,\,+\,\,cflux\,$\times$\,bbody)}. Here, the convolution model \texttt{cflux} was used to calculate the unabsorbed flux for both power law and soft excess components, over the energy range 0.3--8.0\,keV. For the time-averaged spectrum, the ratio of soft excess to primary X-ray integrated flux ($\Phi_{\mathrm{SE}}\textbf{/}\Phi_{\mathrm{PL}}$) is $10^{-12.05}\textbf{/}10^{-10.90} \approx7\%$, with a \texttt{powerlaw} photon index of $\Gamma = 1.98 \pm 0.01$ and a \texttt{blackbody} temperature of $kT = 0.131\pm 0.002 \mathrm{~keV}$.

We then fitted the individual spectrum for each epoch with the above described model for the time-averaged spectrum. Since the \texttt{blackbody} model is used solely to estimate the flux of the soft-excess feature, the $kT$ parameter is not physically meaningful. We allowed $kT$ to remain free throughout the fitting procedure to capture as much flux as possible from the soft excess. In total, our model included four free parameters ---  the photon index $\Gamma$ of the \texttt{powerlaw}, the variable temperature $kT$ of the \texttt{blackbody}, and the fluxes of the two components. We fit spectra from each NICER epoch using the above model and obtain the best-fit parameters for each epoch. 

Figure\,\ref{spectra_mrk359} shows an example spectrum from a single epoch in a high flux state (MJD 59854.50). We find that the data/model ratios are consistent with unity for all individual spectra over the 0.3--8.0\,keV energy range, with a total $C$-statistic of 2770 for 2405\,$\mathrm{dof}$ summed across all epochs for the best-fit models. This suggests that our model is sufficient to estimate the broadband fluxes of the model components. 
The best-fit power law index $\Gamma$ is in the range 1.8--2.1, with a median of 1.9 and a standard deviation of 0.08. The $kT$ is in the range 0.10--0.16\,keV, with a median of 0.13\,keV and a standard deviation of $\sigma=$\,0.02\,keV. 

During our observations, Mrk\,359 is in a state where the soft excess is weaker relative to the primary X-ray continuum than observed in \citealt{2020A&A...640A..99M}. Throughout our observations, the ratio of soft excess to primary X-ray flux, measured over 0.3--8.0\,keV, is in the range $\Phi_{\rm SE}/\Phi_{\rm PL}=2-10\%$. In comparison, the five observations of \citealt{2020A&A...640A..99M} give a higher ratio in the range $\Phi_{\rm SE}/\Phi_{\rm PL}=13-30\%$ over the same energy band. This difference is not simply due to Mrk\,359 being in a lower-flux state overall during our observations. In our NICER monitoring, the total 0.3--8.0\,keV flux spans $0.76-2.02\times10^{-11}\,\mathrm{erg\,cm^{-2}\,s^{-1}}$, which overlaps with and extends above the total flux range of $0.54-1.2\times10^{-11}\,\mathrm{erg\,cm^{-2}\,s^{-1}}$ reported by \citealt{2020A&A...640A..99M}. 

For Mrk\,359, we did not use the \texttt{RELXILLCP} model. In the absence of a detectable Fe K$\alpha$ line in NICER data, the reflection model will use the soft excess to constrain disk parameters. However, the observed soft excess was too weak to constrain the reflection parameters during the fitting process. We thus limit our analysis to the broadband fluxes of the black body and the power law, and the photon index of the power law.

\begin{figure*}[t]
\plotone{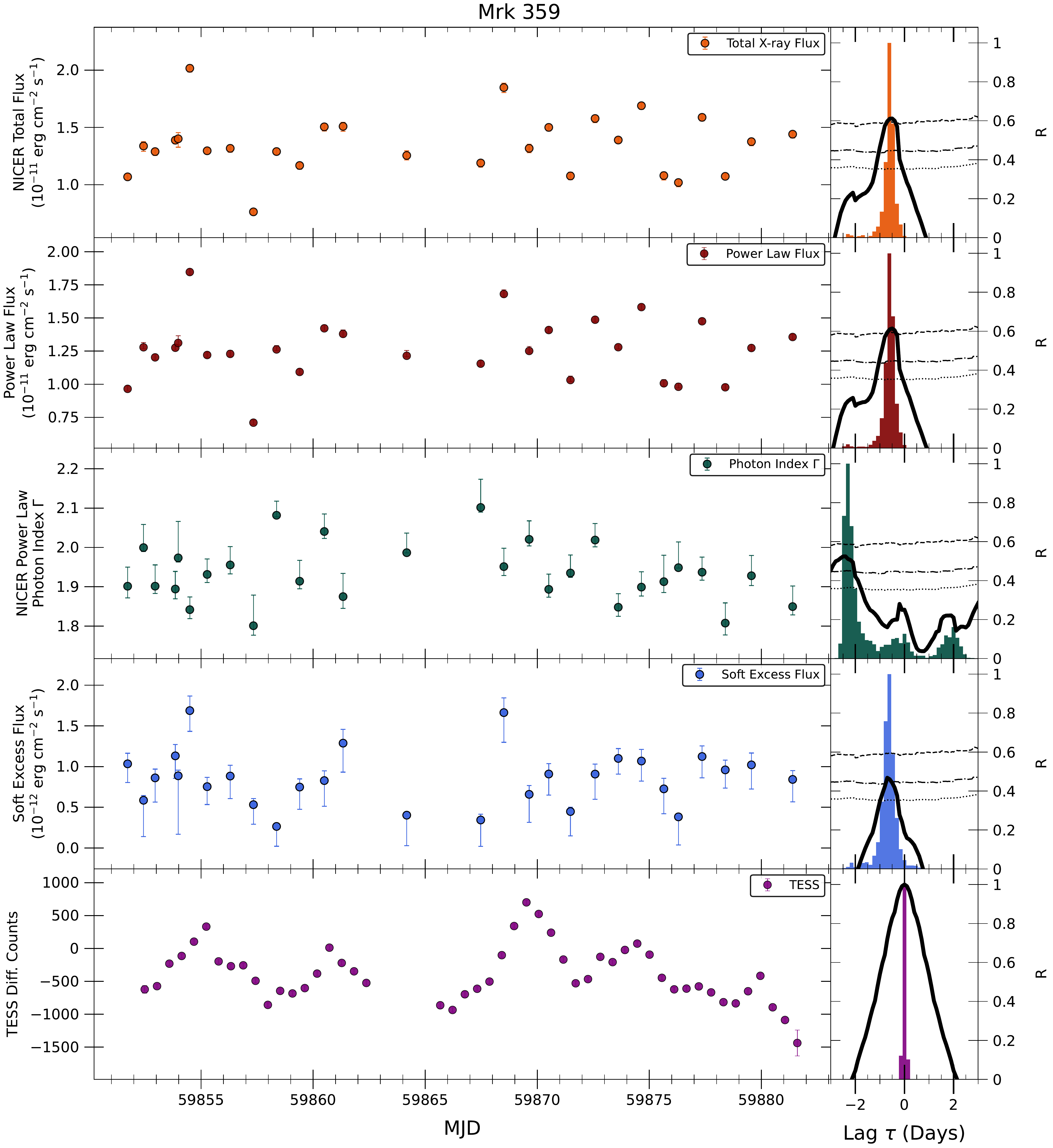}
\caption{\textit{Left:}\,Background-subtracted light curves of Mrk\,359, as observed with NICER and TESS over a 30-day period. The top panel (in orange) shows the total X-ray integrated flux from the best-fit spectral model for each NICER epoch. The second panel (in red color) shows the best-fit integrated flux values from the \texttt{Power Law} component for each NICER epoch. The third panel (in green) shows the photon index of the best-fit power law. The fourth panel (in blue) shows the best-fit flux values for the soft-excess feature, modeled using a \texttt{blackbody} component. The bottom panel (in purple) shows the TESS differential counts binned to 0.5 days, which is half the average cadence of the X-ray data (see \S\ref{subsection:TESS}). \textit{Right:}\,Referenced to the TESS light curve, the right side of the X-ray data panels shows the cross-correlation and lag estimates, as described in \S\ref{sec:cc_analysis}. The bottom panel shows the autocorrelation function of the TESS data (reference light curve). The dotted, dot-dashed, and dashed lines are the 90,\,95,\,and\,99\,per\,cent confidence limits, respectively, calculated using Monte Carlo simulations of light curves assuming no correlation (see Appendix\,\ref{app:simulations}). The probability distributions of the centroid lag values $\tau_\mathrm{cent}$ are also plotted in the right panels. Negative lags indicate that the variability in X-rays occurs first and is then followed by optical variations after a time lag $\tau$. We find no strong correlation between the optical and X-ray light curves. The non-zero $-0.6$\,days lag in the first panel seems to be determined by two high points on the X-ray light curves that precede peaks in the TESS light curve. Refer to Table\,\ref{tab:cc} for the maximum cross correlation coefficients $R_\mathrm{max}$ and lag estimates $\tau_\mathrm{{cent}}$ obtained.
\label{lc_mrk359}}
\end{figure*}

We then obtained the posterior distributions of our free parameters at each epoch using the \texttt{emcee} sampler in \texttt{Python} \citep{emcee}. For Mrk\,359, we used the default stretch move to propose new samples \citep{2010CAMCS...5...65G} using uniform priors. We used a total of 128 walkers, with each walker taking 5000 steps after a 2500-step burn-in period. The total chain length is therefore $\sim$\,$6.4\times10^{5}$. All walkers were initialized at random points in the parameter space, rather than from prior best-fit values. To check for convergence, we inspected the trace plots for good mixing and verified that the chain covers many autocorrelation lengths. The maximum likelihood parameters obtained from the posterior distribution sampled by \texttt{emcee} were consistent with the best-fit parameter values obtained using XSPEC. We give the maximum likelihood parameters, along with the central 68\% confidence intervals, in Table\,\ref{tab:spectral_analysis_mrk359}. We also give the TESS light curve in Table\,\ref{tab:tess_lc_mrk359}. Figure\,\ref{lc_mrk359} shows the resulting background-subtracted light curves of Mrk\,359. 

\setlength{\tabcolsep}{5pt}
\renewcommand{\arraystretch}{1.15}
\begin{deluxetable*}{ccccccc}[]
\tablecaption{Spectral Analysis Results of Mrk\,359.\label{tab:spectral_analysis_mrk359}}
\tablecolumns{7}
\tablehead{
\multicolumn{1}{c}{NICER Epoch} &
\multicolumn{2}{c}{Power Law (Hot Corona)} &
\multicolumn{2}{c}{Black Body (Soft Excess)} &
\multicolumn{1}{c}{Total Source} &
\multicolumn{1}{c}{Fit Statistic} \\
\cmidrule(lr{0.4em}){1-1}
\cmidrule(lr{0.4em}){2-3}
\cmidrule(lr{0.4em}){4-5}
\cmidrule(lr{0.4em}){6-6}
\cmidrule(lr{0.4em}){7-7}
\colhead{MJD} &
\colhead{$\Phi_{\mathrm{PL}}$} &
\colhead{$\Gamma$} &
\colhead{$\Phi_{\mathrm{BB}}$} &
\colhead{$kT$} &
\colhead{$\Phi_{\rm t} = \Phi_{\mathrm{PL}} + \Phi_{\mathrm{BB}}$} &
\colhead{} \\
\colhead{} &
\colhead{$[10^{-11}\ \mathrm{erg\,cm^{-2}\,s^{-1}}]$} &
\colhead{} &
\colhead{$[10^{-12}\ \mathrm{erg\,cm^{-2}\,s^{-1}}]$} &
\colhead{[keV]} &
\colhead{$[10^{-11}\ \mathrm{erg\,cm^{-2}\,s^{-1}}]$} &
\colhead{C-stat (\rm dof)}
}
\startdata
$59851.72$ & $0.97^{+0.02}_{-0.02}$ & $1.90^{+0.05}_{-0.03}$ & $1.03^{+0.13}_{-0.23}$ & $0.14^{+0.01}_{-0.01}$ & $1.07^{+0.03}_{-0.03}$ & $81\ (82)$ \\
$59852.43$ & $1.28^{+0.04}_{-0.01}$ & $2.00^{+0.06}_{-0.01}$ & $0.59^{+0.06}_{-0.45}$ & $0.13^{+0.03}_{-0.01}$ & $1.34^{+0.04}_{-0.05}$ & $90\ (86)$ \\
$59852.95$ & $1.20^{+0.02}_{-0.02}$ & $1.90^{+0.05}_{-0.02}$ & $0.86^{+0.11}_{-0.30}$ & $0.13^{+0.01}_{-0.01}$ & $1.29^{+0.03}_{-0.03}$ & $95\ (85)$ \\
$59853.85$ & $1.28^{+0.03}_{-0.02}$ & $1.89^{+0.04}_{-0.02}$ & $1.13^{+0.14}_{-0.28}$ & $0.13^{+0.01}_{-0.01}$ & $1.39^{+0.03}_{-0.03}$ & $100\ (84)$ \\
$59854.98$ & $1.31^{+0.05}_{-0.01}$ & $1.97^{+0.09}_{-0.01}$ & $0.89^{+0.07}_{-0.72}$ & $0.13^{+0.03}_{-0.01}$ & $1.40^{+0.05}_{-0.07}$ & $84\ (77)$ \\
$59854.50$ & $1.85^{+0.02}_{-0.02}$ & $1.84^{+0.03}_{-0.02}$ & $1.69^{+0.18}_{-0.25}$ & $0.11^{+0.00}_{-0.00}$ & $2.02^{+0.03}_{-0.03}$ & $61\ (88)$ \\
$\cdots$ & $\cdots$ & $\cdots$ & $\cdots$ & $\cdots$ & $\cdots$ & $\cdots$ \\
Time-averaged & $1.261^{+0.004}_{-0.004}$ & $1.98^{+0.01}_{-0.01}$ & $0.89^{+0.04}_{-0.04}$ & $0.131^{+0.002}_{-0.002}$ & $1.349^{+0.006}_{-0.006}$ & $\chi^{2}/\mathrm{dof} =134/114$\enddata
\tablecomments{The maximum likelihood parameters from spectral analysis of each NICER epoch for Mrk\,359, as obtained using the \texttt{emcee} sampler. The errors are central $68\%$ confidence intervals from the posterior distribution. This table is available in its entirety in the online journal.
}
\end{deluxetable*}

\begin{deluxetable}{cc}
\tablewidth{0pt}
\tablecaption{TESS Light Curve of Mrk\,359\label{tab:tess_lc_mrk359}}
\tablehead{
\colhead{MJD} &
\colhead{Differential Counts ($\pm 1\sigma$ uncertainty)}
}
\startdata
59852.49 & $-621.96 \pm 57.92$ \\
59853.03 & $-573.24 \pm 33.17$ \\
59853.58 & $-231.50 \pm 33.57$ \\
59854.13 & $-114.30 \pm 33.49$ \\
59854.68 & $103.18 \pm 33.58$ \\
$\cdots$ & $\cdots$ \\
\enddata
\tablecomments{This table is available in its entirety in the online journal.}
\end{deluxetable}

\subsection{NGC\,4051} \label{subsec:NGC4051}

\begin{figure*}[t]
\centering
\includegraphics[width=\textwidth]{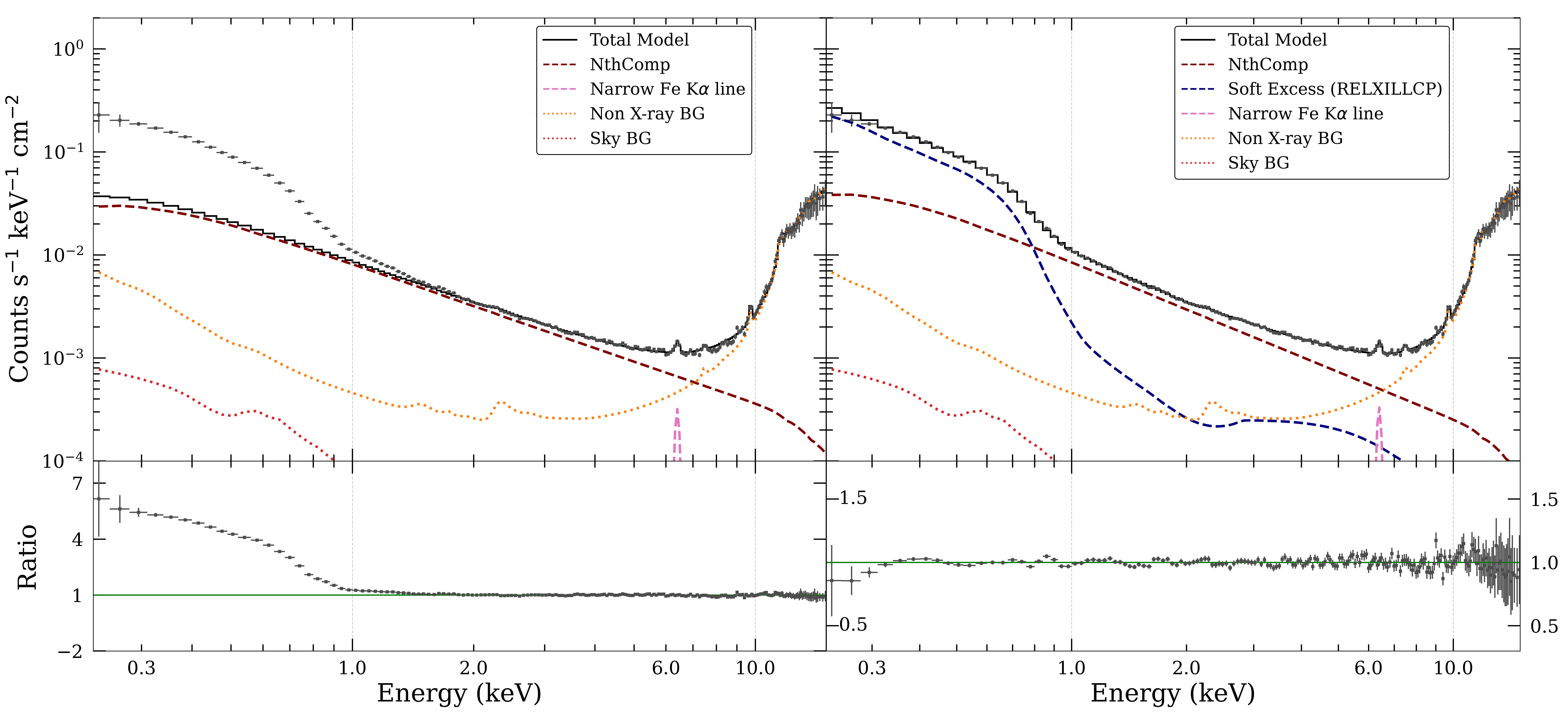}
\caption{Time-averaged X-ray spectrum of all NICER epochs for NGC\,4051, similar to Figure\,\ref{average_spectrum_mrk359}. Using the \texttt{RELXILLCP} model for soft excess gives a $\chi^{2}/\mathrm{dof} = 129/113 = 1.14$. For the time-averaged spectrum, the ratio $\Phi_{\mathrm{SE}}/\Phi_{\mathrm{PL}}$ in the energy range $0.3$--$8.0~\mathrm{keV}$ is $10^{-10.89}/10^{-10.69} \approx 60\%$, with a photon index $\Gamma = 1.55^{+0.05}_{-0.10}$.
\label{average_spectrum_ngc4051}
}
\end{figure*}

Similar to Mrk\,359, we first fit a single absorbed power-law model over the 2--15 keV energy range. The hydrogen column density $N_{\rm H}$ for \texttt{tbabs} was set to $1.19\times 10^{20} \mathrm{~cm^{-2}}$ \citep{2016A&A...594A.116H}. The single absorbed power-law model fitted to the time-averaged spectrum gives a power law photon index $\Gamma = 1.32 \pm 0.02$, with a $\chi^{2}/\mathrm{dof} = 131/80$. The time-averaged spectrum shows iron emission-line residuals, see Figure\,\ref{average_spectrum_ngc4051}, which we further fit with a narrow \texttt{zgauss} component ($\sigma = 0$) at a redshift of $z = 0.0023$. The addition of a \texttt{zgauss} component improves the $\chi^{2}$ by $60$ for $2$ $\mathrm{dof}$, with a total $\chi^{2}/\mathrm{dof} = 71/78$ in the 2--15$\mathrm{~keV}$ energy range. The resulting photon index $\Gamma = 1.32 \pm 0.01$ is consistent with the previous fit. Extrapolating the best-fit hard X-ray model to lower energies reveals strong positive residuals below $\sim2\,\mathrm{keV}$, confirming the presence of a soft excess, as shown in Figure\,\ref{average_spectrum_ngc4051}.

We found that the phenomenological \texttt{blackbody} model is insufficient to fit the strong soft excess feature in NGC\,4051. A \texttt{powerlaw\,+\,blackbody\,+\,zgauss} model with the best-fit value of $kT = 0.1 \,\rm keV$ gave a poor fit to the time-averaged spectrum with a $\chi^{2}/\mathrm{dof} = 500/115$. Replacing the \texttt{blackbody} component with \texttt{Nthcomp} also provides a poor fit, with $\chi^{2}/\mathrm{dof} = 484/116$, with $kT_{e} = 0.1 \,\rm keV$ and an extreme $\Gamma_{\rm wc} = 1.0$, compared to the much steeper photon indices of $\Gamma_{\rm wc}\sim2.0-3.0$ typically found for other AGN and suggestive of warm-coronal emission \citep{2018A&A...611A..59P}.
Instead, we employed the high-density disk reflection model \texttt{RELXILLCP} to fit the soft excess. We now model the hard X-ray primary continuum using the thermally Comptonized continuum model \texttt{Nthcomp} \citep{1996MNRAS.283..193Z,1999MNRAS.309..561Z} instead of a \texttt{powerlaw} model, to remain consistent with the internal primary continuum model used in \texttt{RELXILLCP}. We used \texttt{RELXILLCP} solely to model the reflection component, fixing the reflection fraction parameter to $-1$ so that the model returns only the reflected spectrum. In XSPEC terminology, our final model for the time-averaged spectrum is \texttt{tbabs} $\times$ (\texttt{cflux} $\times$ \texttt{Nthcomp} + \texttt{cflux} $\times$ \texttt{relxillCp} + \texttt{cflux} $\times$ \texttt{zgauss}).

The model \texttt{Nthcomp} has four key parameters: the power-law photon index $\Gamma$, the electron temperature $kT_{\rm e}$, which determines the high-energy rollover; the blackbody seed photon temperature $kT_{\rm bb}$, which decides the low-energy rollover; and the normalization. We fixed $kT_{\rm bb}=0.01\,\mathrm{keV}$ to match the seed-photon temperature used in \texttt{RELXILLCP}. Since NICER's energy range ($<$\,15\,keV) lies well below the expected high-energy rollover (typically $100 - 300 \mathrm{~keV}$), we fixed $kT_{\rm e}$ at $100 \mathrm{~keV}$ throughout the fitting process. 

The parameters to model the primary source in \texttt{RELXILLCP} were tied to those of the \texttt{Nthcomp} component. The model parameters and their allowed ranges are listed in Table~\ref{tab:model_params_ngc4051}. We modeled the reflection spectrum assuming an accretion disk with an outer radius of $R_{\rm out} = 1000\,R_{\rm g}$. For the time-averaged spectrum, we leave the inclination angle with respect to the disk normal as a free parameter. We fixed the black hole spin to the maximal value of $a = 0.998$, as the broad Fe-line used to model the spin is not significantly detected in NICER spectra. A reflection model with $a=0$ also did not adequately reproduce the strong and broad soft excess observed and left positive residuals below 2\,keV. We therefore chose the maximal value of spin, $a = 0.998$, to allow the reflection model to fit the observed soft excess. This version of \texttt{relxill} does not assume a lamp post geometry, and instead uses an empirical broken power law emissivity profile as a function of radius to describe the flux irradiating the accretion disk. Below the break radius $R_{\rm br}$, the emissivity follows $r^{-\mathrm{Index1}}$, and beyond $R_{\rm br}$ it falls off as $r^{-\mathrm{Index2}}$. The break radius distinguishes the inner and outer regions of the disk, and was fixed at $R_{\rm br} = 6\,R_{\rm g}$ in our model. \citealt{2012MNRAS.424.1284W} computed emissivity profiles for AGN and found that within the inner region, the emissivity falls steeply (with indices between 6 and 8), while in the outer region of the disk it flattens to an index of 3. Consequently, we let $\mathrm{Index1}$ vary freely and fixed $\mathrm{Index2} = 3$ to model the disk emissivity profile of our source. The ionization ($\xi$), density ($n$), and iron abundance ($A_{\rm Fe}$) of the accretion disk were kept as free parameters.

\begin{deluxetable}{ccc}[]
\tablecaption{Source Model Parameters for NGC~4051 \label{tab:model_params_ngc4051}}
\tablecolumns{3}
\tablewidth{0pt}
\tablehead{
\colhead{Model} &
\colhead{Parameter [unit]} &
\colhead{Allowed Range (Priors)}
}
\startdata
tbabs &
$N_{\rm H}~[10^{22}\,\mathrm{cm^{-2}}]$ &
$0.0119$ \\
\hline
cflux &
$ \log(\Phi_{\mathrm{PL}})~[\mathrm{erg~cm^{-2}~s^{-1}}]$ &
$[-12,-10.0]$ \\ 
NthComp &
Photon Index $\Gamma$ &
$[1,3]$ \\
\hline
cflux &
$ \log(\Phi_{\mathrm{SE}})~[\mathrm{erg~cm^{-2}~s^{-1}}]$ &
$[-12.5,-10.0]$ \tablenotemark{a} \\
relxillCp &
Inclination [$^\circ$] &
$[10,87]$ \\
relxillCp &
Spin ($a$) &
$0.998$ \\
relxillCp &
Inner Emissivity Index $q_{1}$ &
$[3,10]$ \\
relxillCp &
Emissivity Break Radius $R_{\rm br}$ [$R_{\rm g}$] &
$6$ \\
relxillCp &
Outer Emissivity Index $q_{2}$ &
$3$ \\
relxillCp &
$\log \xi~[\mathrm{erg~cm~s^{-1}}]$ &
$[0,4.7]$ \\
relxillCp &
$\log n~[\mathrm{cm^{-3}}]$ &
$[15,20]$ \\
relxillCp &
Iron Abundance $A_{\mathrm{Fe}}$ [$A_\odot$] &
$[0.5,10]$ \\
\enddata
\tablenotetext{a}{During the MCMC sampling, we imposed a prior requiring the flux of the reflection component \texttt{relxillCp} to be less than the flux of the \texttt{NthComp} component. This prior excludes solutions in which the reflection component alone models the entire spectrum, while the primary \texttt{NthComp} component is suppressed. This condition ensures that the \texttt{NthComp} component accounts for the bulk of the primary X-ray continuum, while the \texttt{relxillCp} contributes predominantly to the soft excess feature.}
\end{deluxetable}


\figsetstart
\figsetnum{5}
\figsettitle{Full set of NICER spectra and fitted models of NGC\,4051}

\figsetend

\begin{figure*}[t]
\centering
\includegraphics[width=\textwidth]{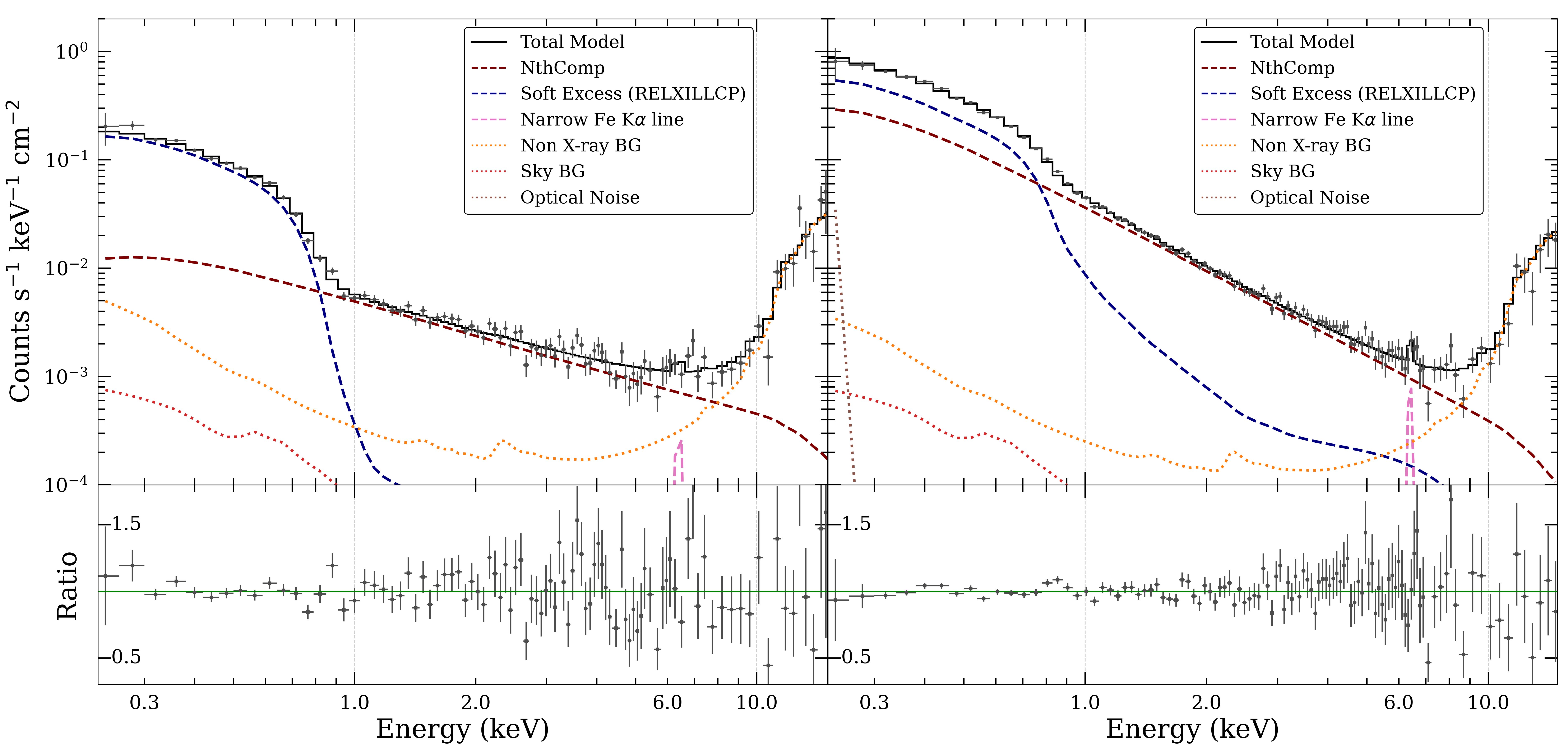}
\caption{X-ray spectrum from two NICER epochs of NGC\,4051, MJD 59648.21 (left) and MJD 59653.89 (right), as observed in low (left) flux and high (right) flux states, respectively. For the low flux state, the ratio $\Phi_{\mathrm{SE}}/\Phi_{\mathrm{PL}}$ in the energy range $0.3$--$8.0~\mathrm{keV}$ is $10^{-10.93}/10^{-10.71} \approx 60\%$ but with a hard photon index $\Gamma = 1.1^{+0.03}_{-0.03}$. For the high flux state, $\Phi_{\mathrm{SE}}/\Phi_{\mathrm{PL}} = 10^{-10.42}/10^{-10.18} \approx 60\%$, with a soft photon index $\Gamma = 1.97^{+0.01}_{-0.02}$. The high flux state corresponds to an X-ray flare as reported in \citealt{2025arXiv250113049R}. The spectra and fitted models for all epochs are available as a figure set in the online journal.}
\label{single_epochs_spectrum_ngc4051}
\end{figure*}


\begin{figure*}[t]
\plotone{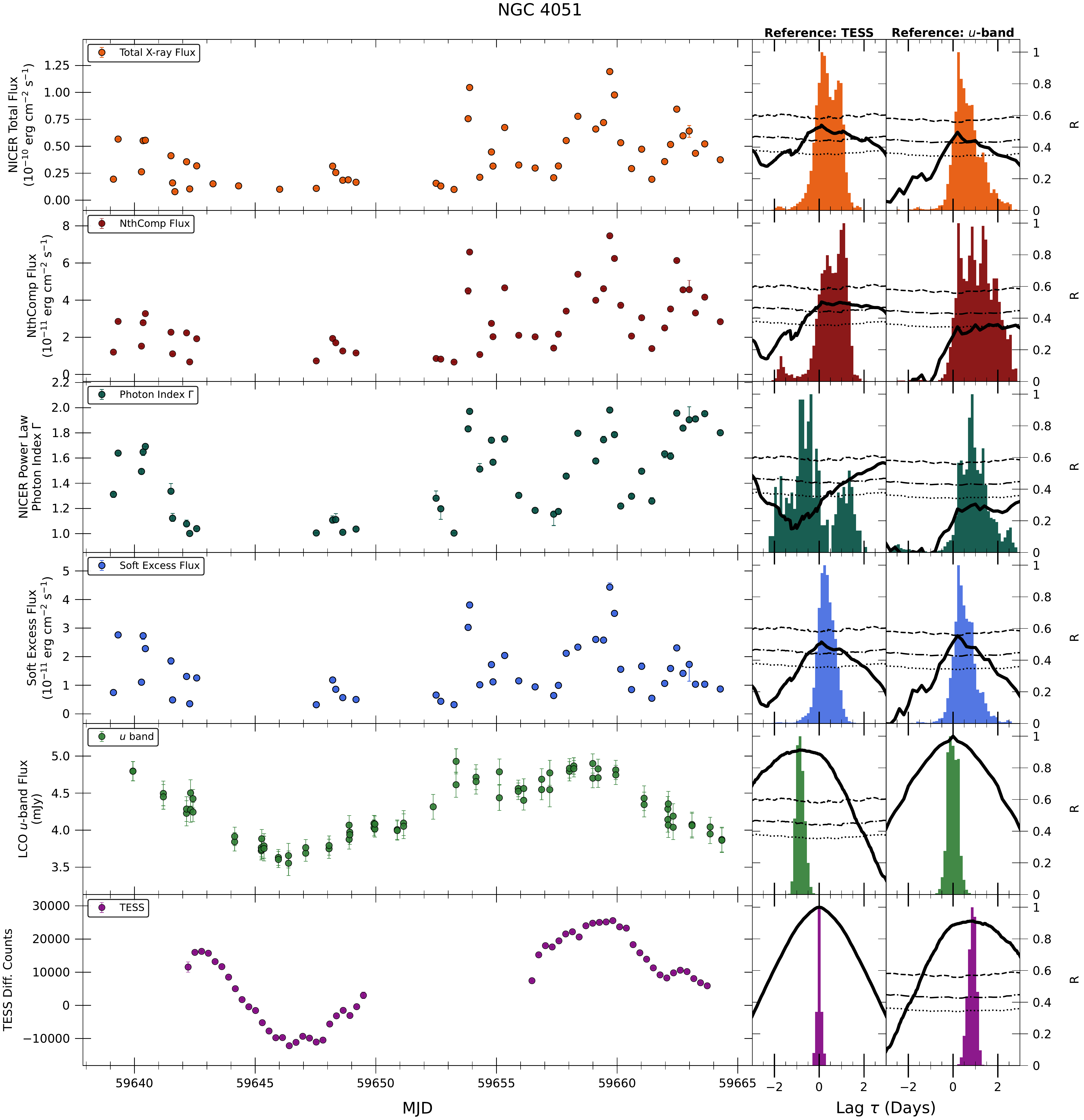}
\caption{Same as Figure\,\ref{lc_mrk359}, but for NGC\,4051, and with $u$\,band light curve observed with LCO and as produced in \citealt{2026MNRAS.548ag642M,doi:10.5281/zenodo.17242993}. For NICER observations between MJD 59643-59648, the individual components were not constrained; therefore, we report only the total flux. TESS data is binned to 0.28 days (refer \S\ref{subsection:TESS}). The second and third columns show the cross-correlation functions and centroid distributions of the time lags obtained using the TESS and $u$-band light curves as the reference, respectively. We find no strong correlation between the optical and X-ray light curves. The two cross-correlation analyses give consistent results.
\label{lc_ngc4051}}
\end{figure*}

\setlength{\tabcolsep}{3.5pt}
\renewcommand{\arraystretch}{1.15}
\begin{deluxetable*}{cccccccccccc}[]
\tablecaption{Spectral Analysis Results of NGC~4051.\label{tab:spectral_analysis_ngc4051}}
\tablecolumns{12}
\tablehead{
\multicolumn{1}{c}{NICER Epoch} &
\multicolumn{2}{c}{NthComp} &
\multicolumn{6}{c}{RELXILLCP (Soft Excess)} &
\multicolumn{1}{c}{zgauss \tablenotemark{a}} &
\multicolumn{1}{c}{Total Source} &
\multicolumn{1}{c}{Statistic} \\
\cmidrule(lr{0.4em}){1-1}
\cmidrule(lr{0.4em}){2-3}
\cmidrule(lr{0.4em}){4-9}
\cmidrule(lr{0.4em}){10-10}
\cmidrule(lr{0.4em}){11-11}
\cmidrule(lr{0.4em}){12-12}
\colhead{MJD} &
\colhead{$\Phi_{\mathrm{NthComp}}$} &
\colhead{$\Gamma$} &
\colhead{$\Phi_{\mathrm{REL}}$} &
\colhead{$\log n$} &
\colhead{$q_{1}$} &
\colhead{$\mathrm{log}\,\xi$} &
\colhead{$i$} &
\colhead{$A_{\mathrm{Fe}}$} &
\colhead{$\Phi_{\mathrm{Fe}}$} &
\colhead{$\Phi_{\mathrm{t}}$} &
\colhead{} \\
\colhead{} &
\colhead{$[10^{-11}]$} &
\colhead{} &
\colhead{$[10^{-11}]$} &
\colhead{$[\mathrm{cm^{-3}}]$} &
\colhead{} &
\colhead{$[\mathrm{erg\,cm\,s^{-1}}]$} &
\colhead{[deg]} &
\colhead{[$\mathrm{Solar}$]} &
\colhead{$[10^{-13}]$} &
\colhead{$[10^{-11}]$} &
\colhead{C-stat ($\rm dof$)}
}
\startdata
$59639.13$ &
$1.20^{+0.01}_{-0.03}$ &
$1.31^{+0.01}_{-0.02}$ &
$0.75^{+0.04}_{-0.01}$ &
$19.95^{+0.00}_{-0.20}$ &
$8.3^{+0.2}_{-0.1}$ &
$2.15^{+0.04}_{-0.01}$ &
$66.5^{+0.3}_{-0.3}$ &
$2.6^{+0.2}_{-0.1}$ &
$\leq$\,1.2 &
$1.95^{+0.04}_{-0.03}$ &
$190\,(88)$ \\
$59639.32$ &
$2.85^{+0.02}_{-0.04}$ &
$1.64^{+0.00}_{-0.02}$ &
$2.76^{+0.04}_{-0.00}$ &
$19.98^{+0.01}_{-0.10}$ &
$7.9^{+0.1}_{-0.2}$ &
$1.79^{+0.03}_{-0.01}$ &
$\cdots$ &
$\cdots$ &
$\leq$\,13.7 &
$5.66^{+0.06}_{-0.06}$ &
$120\,(91)$ \\
$59640.29$ &
$1.52^{+0.02}_{-0.04}$ &
$1.49^{+0.02}_{-0.02}$ &
$1.11^{+0.02}_{-0.02}$ &
$19.94^{+0.02}_{-0.16}$ &
$8.8^{+0.1}_{-0.3}$ &
$1.95^{+0.04}_{-0.02}$ &
$\cdots$ &
$\cdots$ &
$\leq$\,3.2
&
$2.63^{+0.03}_{-0.05}$ &
$119\,(84)$ \\
$59640.36$ &
$2.78^{+0.12}_{-0.00}$ &
$1.65^{+0.01}_{-0.03}$ &
$2.73^{+0.02}_{-0.13}$ &
$19.62^{+0.31}_{-0.10}$ &
$8.5^{+0.2}_{-0.6}$ &
$1.88^{+0.01}_{-0.07}$ &
$\cdots$ &
$\cdots$ &
$\leq$\,2.5 &
$5.52^{+0.12}_{-0.13}$ &
$102\,(80)$ \\
$59640.45$ &
$3.27^{+0.03}_{-0.03}$ &
$1.69^{+0.00}_{-0.03}$ &
$2.28^{+0.04}_{-0.02}$ &
$19.91^{+0.07}_{-0.02}$ &
$7.7^{+0.3}_{-0.1}$ &
$1.74^{+0.05}_{-0.01}$ &
$\cdots$ &
$\cdots$ &
$\leq$\,4.9 &
$5.56^{+0.05}_{-0.04}$ &
$122\,(96)$ \\
$59641.51$ &
$2.27^{+0.02}_{-0.10}$ &
$1.34^{+0.06}_{-0.01}$ &
$1.85^{+0.02}_{-0.12}$ &
$19.70^{+0.23}_{-0.01}$ &
$8.6^{+0.4}_{-0.1}$ &
$2.18^{+0.01}_{-0.05}$ &
$\cdots$ &
$\cdots$ &
$\leq$\,1.6 &
$4.12^{+0.03}_{-0.15}$ &
$118\,(79)$ \\
$\cdots$ & $\cdots$ & $\cdots$ & $\cdots$ & $\cdots$ & $\cdots$ & $\cdots$ & $\cdots$ & $\cdots$ & $\cdots$ & $\cdots$ & $\cdots$ \\
$\mathrm{Time\mbox{-}averaged}$ &
$2.05^{+0.01}_{-0.14}$ &
$1.55^{+0.05}_{-0.01}$ &
$1.29^{+0.14}_{-0.01}$ &
$19.5^{+0.1}_{-0.3}$ &
$9.97^{+0.03}_{-4.08}$ &
$1.93^{+0.15}_{-0.01}$ &
$66^{+1}_{-35}$ &
$7.0^{+2.9}_{-0.2}$ &
$2.4^{+0.3}_{-0.3}$ &
$3.37^{+0.14}_{-0.14}$ &
$\chi^{2}/\mathrm{dof} = 129/113$ \vspace{5pt}\\
\enddata
\tablenotetext{a}{$E = 6.4\,\rm keV$, $\sigma = 0 \,\rm keV$. Upper limits are given at $3\sigma$ confidence level.}
\tablecomments{Same as Table~\ref{tab:spectral_analysis_mrk359} but for NGC~4051.
Fluxes are reported in units of $\mathrm{erg\,cm^{-2}\,s^{-1}}$. 
This table is available in its entirety in the online journal.}
\end{deluxetable*}

\begin{deluxetable}{cc}
\tablewidth{0pt}
\tablecaption{TESS Light Curve of NGC\,4051\label{tab:tess_lc_NGC4051}}
\tablehead{
\colhead{MJD} &
\colhead{Differential Counts ($\pm 1\sigma$ uncertainty)}
}
\startdata
59642.22 & $11546.75 \pm 1554.50$ \\
59642.50 & $15968.12 \pm 241.38$ \\
59642.78 & $16275.49 \pm 231.89$ \\
59643.06 & $15712.17 \pm 231.75$ \\
59643.34 & $13193.54 \pm 248.15$ \\
$\cdots$ & $\cdots$ \\
\enddata
\tablecomments{This table is available in its entirety in the online journal.}
\end{deluxetable}

For the time-averaged spectrum, the above model gives a best-fit statistic of $\chi^{2}/\mathrm{dof} = 129/113$ and with small residuals across the passband in the energy range 0.3--8.0 \, keV, as shown in the ratio plot in Figure\,\ref{average_spectrum_ngc4051}. This suggests that the above described model can be used to estimate the broadband fluxes of the soft-excess and primary X-ray components in the 0.3-8.0\,keV range for individual epoch spectra. The ratio of soft excess to primary X-ray integrated flux ($\Phi_{\mathrm{SE}}\textbf{/}\Phi_{\mathrm{PL}}$) is $10^{-10.89}\textbf{/}10^{-10.69} \approx60\%$ with a photon index $\Gamma = 1.55^{+0.05}_{-0.01}$. The best-fit accretion disk density parameter is $\log n = 19.5^{+0.1}_{-0.3} $, the inclination is $i = 66^{+1}_{-35} \deg$, and iron abundance is $A_{\rm Fe}=7.0^{+2.9}_{-0.2}$ times the Solar value for the time-averaged spectrum.

We then model individual NICER epoch using the above described model for the time-averaged spectrum. The inclination angle and the iron abundance of the disk are not expected to exhibit strong variability over the time baseline of the observations. We therefore tied these two parameters and jointly fitted all individual epoch spectra. For NICER observations between MJD 59643-59648, the individual components were not constrained due to a combination of low source flux and high background level; therefore, we remove these 5 spectra from the joint fit. We find the best-fit model for the joint fit using the Levenberg-Marquardt C-stat minimization algorithm in \texttt{XSPEC}.

We find consistent ratio plots across all epochs for the best joint-fit model, with a $C-\rm statistic = 5413$ for 4024\,$\rm dof$. 
We then sample from the posterior distributions of free source parameters of the joint fit using \texttt{emcee}, in a similar manner to that used for Mrk\,359. However, since the model for NGC\,4051 is more complex and the joint fit has $324$ free source parameters, we used a weighted mixture of two moves to propose new samples, a differential evolution (DE) move with $80\%$ probability and a snooker move using differential evolution (DESnooker) with $20\%$ probability \citep{2014ApJS..210...11N,10.1007/s11222-008-9104-9}. The priors for all parameters were uniform but bounded within the allowed ranges specified in Table~\ref{tab:model_params_ngc4051}. We used a total of 648 walkers, with each walker taking 15000 steps and a burn-in period of 3000 steps, resulting in a total chain length of $\sim$$1\times 10^{7}$. All walkers were initialized using correlated realizations generated from the covariance matrix of the best joint-fit XSPEC solution using \texttt{simpars} command in \texttt{XSPEC}. We checked for convergence by visually inspecting the trace plots to assess good mixing and verified that the chains covered many autocorrelation lengths. We find consistent results from the \texttt{emcee} samples and \texttt{XSPEC} joint fit.

We give the maximum likelihood parameters, along with the central 68\% confidence intervals in Table\,\ref{tab:spectral_analysis_ngc4051}. Figure\,\ref{single_epochs_spectrum_ngc4051} shows example spectra from two epochs, corresponding to low- and high-flux states and Figure\,\ref{corner_NGC4051} also shows the posterior distribution of this high-flux state MJD 59653.89. The high-flux state corresponds to the X-ray flare first reported by \citealt{2025arXiv250113049R}, during which the $0.3-8.0\mathrm{\,keV}$ flux increased by a factor of $\sim$10 and returned to a similar level within one day (MJD 59653). The ratio of soft excess to primary X-ray flux is in the range of 0.25--0.98, indicating that the soft excess is a significant fraction of the observed X-ray emission throughout the monitoring period. Table\,\ref{tab:tess_lc_NGC4051} gives the TESS differential counts and Figure\,\ref{lc_ngc4051} shows both the NICER and TESS light curves along with the cross-correlation results described in \S\,\ref{sec:cc_analysis}.

The median value for the density of the accretion disk is $10^{19.8} \mathrm{~cm^{-3}}$ with a standard deviation of 0.3\,dex. This is about $5$ orders of magnitude greater than the commonly assumed density value of $10^{15} \mathrm{~cm^{-3}}$ adopted in X-ray reflection models \citep{2016MNRAS.462..751G}, and is close to the limit of the parameter space covered by the models ($10^{20} \mathrm{~cm^{-3}}$). The upper limit on the disk density in the \texttt{relxill} model is due to a lack of atomic data needed to calculate reflection from high-density plasmas \citep{2016MNRAS.462..751G}. This raises a concern that the resulting fits move to unphysical parts of parameter space, as other parameters try to compensate for the limit on the density.

The density is weakly correlated with the overall flux level ($\Phi_{\rm NthComp}$ and $\Phi_{\rm REL}$) and weakly anti-correlated with the ionization parameter, so these are the main parameters we would expect to be most affected by the model limit on the density (other parameters may also be affected through correlations with these parameters, such as photon index). Although an independent estimate of these parameters is needed to check the fitted values, the crux of our experiment relies on measuring the variability of the soft excess component. Even if the fitted values are biased relative to the physical values, our modeling is adequate to capture variations in the soft excess and extract a clean light curve of this component.  We note that the high density values are consistent with studies of other AGN, which have found that that high-density accretion disks models are required to fit the soft excess, and we do not require ad hoc spectral components to fit the soft excess (c.f., \citealt{2022MNRAS.516.5027O}, \citealt{2019MNRAS.489.3436J},\citealt{2022MNRAS.513.4361M},\citealt{2024MNRAS.534..608M},\citealt{2025arXiv250115380M}). The fitted high density model therefore is a reasonable description of the data, and suggests a consistent comparison with previous work, even if the fitted parameter is biased from the physical value.

The maximum likelihood model has an inclination of $i=66.5^{+0.3}_{-0.3} \deg$, consistent with the time-averaged fit, whereas a lower iron-abundance is inferred, with $A_{\rm Fe}= 2.6^{+0.2}_{-0.1}$ rather than $A{\rm Fe}= 7.0^{+2.9}_{-0.2}$. The higher abundance inferred from the time-averaged spectrum may result from averaging over epoch-dependent spectral-shape variability and detection of Fe\,K$\alpha$ line. The \texttt{RELXILLCP} model required a steep emissivity profile at the inner parts of the disk to fit the strong soft excess. The median value for $\mathrm{Index1}$ is $9$, highlighting that the reflection profile requires strong irradiation at the inner regions of the disk that is proportional to $r^{-9}$ compared to $r^{-3}$ expected for a flat space-time. An $r^{-3}$ emissivity profile cannot model the strong soft excess in our data. The ionization parameter ranges from $\log \xi \sim 1.4$--$2.9 \rm\,erg\,cm\,s^{-1}$, with a median value of $\log \xi = 2.2\rm\,erg\,cm\,s^{-1}$. We, however, note that the disk parameters inferred for NGC\,4051 can be partially degenerate due to limitations in the reflection modeling and low S/N in the Fe K bandpass.

\begin{figure}[]
  \centering
  \includegraphics[width=\columnwidth]{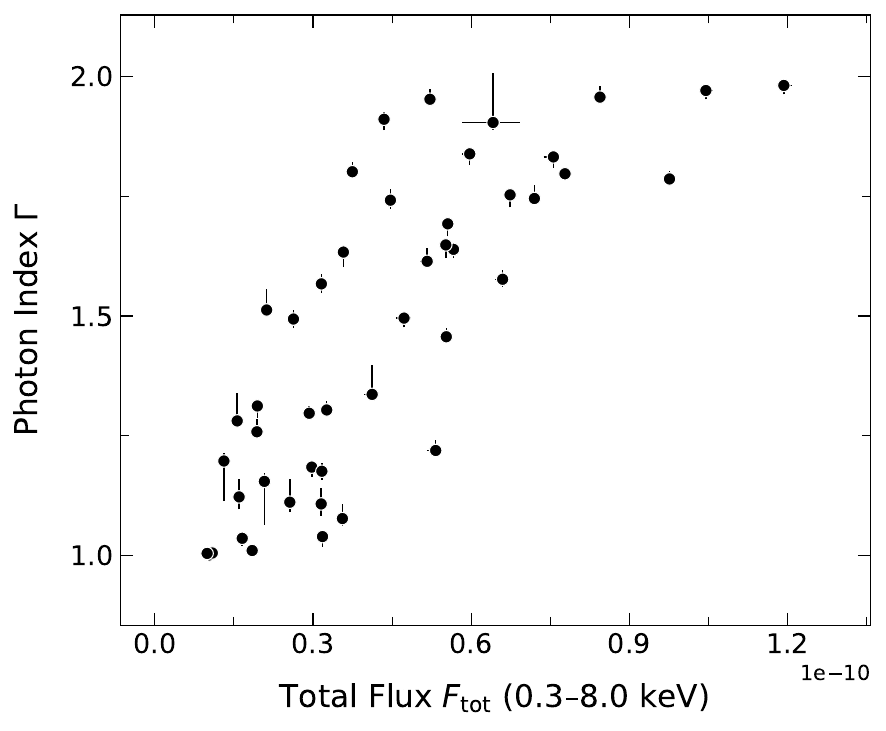} 
  \caption{Total X-ray flux versus the photon index $\Gamma$ from the time-resolved spectroscopic fits of NGC\,4051 NICER observations. Although no extremely tight correlation exists, a positive trend is evident, consistent with the softer-when-brighter behavior seen in other AGN \citep{2009MNRAS.399.1597S}.}
  \label{flux_gamma_ngc4051}
\end{figure}

We show the correlation between photon index $\Gamma$ for the \texttt{Nthcomp} component and the total 0.3--8.0\,keV flux in Figure\,\ref{flux_gamma_ngc4051}. The photon index $\Gamma$ covers a range of 1.0--2.0 with a median of 1.49 and a standard deviation of 0.32. The flatter primary X-ray slopes are associated with the low flux states, while the higher flux states show a softer spectrum, as has been seen for this source before \citep{2006MNRAS.368..903P, 2009MNRAS.399.1597S} and similar to a softer-when-brighter trend seen in other AGN \citep{2009MNRAS.399.1597S}. 

One ambiguity for NGC\,4051 is the very hard spectrum ($\Gamma< 1.5$) during periods of low flux. It is possible that absorption removes flux at low energies, while the overall NICER spectrum can still be modeled by a shallow power law because of because of NICER's limited bandpass. To test whether the flux removed at lower energies modifies the variability shape of the X-ray light curves shown in Figure\,\ref{lc_ngc4051}, we also perform time-resolved spectroscopic fits with an absorption model. We assumed a \texttt{powerlaw} with a photon index $\Gamma=2.0$ (fixed) and a \texttt{blackbody} with $kT=0.12\rm\,keV$ (fixed), both modified by two variable ionized absorbers in the nucleus modelled using \texttt{zxipcf}. The fluxes of both additive model components were left free. In \texttt{XSPEC} terminology, the spectral model is expressed as: \texttt{tbabs}\,$\times$\allowbreak\,\texttt{zxipcf}\,$\times$\allowbreak\,\texttt{zxipcf}\,$\times$\allowbreak\,(\texttt{cflux}\,$\times$\allowbreak\,\texttt{powerlaw}\,+\,\texttt{cflux}\,$\times$\allowbreak\,\texttt{bb}). The model \texttt{zxipcf} uses a grid of XSTAR photoionized absorption models (calculated assuming a microturbulent velocity of $200\,\rm km\,s^{-1}$) for the absorption. The column density $N_{\rm H}$ and the ionization parameter $\log \xi$ where $\xi = L/nr^{2}$ of the \texttt{zxipcf} were left free for fitting. We fixed the covering fraction of \texttt{zxipcf} to 1.0, to reduce degeneracies with the column density. We also fixed the redshift to the source value of $z = 0.0023$, which remained unconstrained if left free. For certain epochs, narrow ($\sigma=0$) gaussian lines at energies $\sim$\,$0.9\,\rm keV$ and $\sim$\,$0.6\,\rm keV$ were also required to improve the fit statistic. We fit spectra from each NICER epoch using the absorption model and obtain the best-fit parameters for each epoch in XSPEC. Over the energy range 0.3-8.0\,keV for which integrated flux is calculated for the additive model components, we found the total $\Sigma \chi^{2}/\mathrm{dof}$\,$=$\,4854/3540\,$=$\,1.37, summed across epochs where the model was constrained. We show the X-ray light curves obtained using the absorption spectral model in Figure\,\ref{lc2_ngc4051}. It is visually evident that the variability shape of the X-ray light curves remain the same, and are only shifted in overall normalization. We confirm this by cross correlating the X-ray light curves obtained using the two different spectral models, and obtain a cross correlation coefficient of $R = 0.95,0.95,0.9$ for total, power-law, and soft excess fluxes. Thus, the two spectral models give the same light curve shape. Consequently, this result implies that the strong X-ray flux variability of NGC\,4051 is dominated by intrinsic variations in the continuum flux rather than due to any variable absorption that modifies the power-law shape. Thus, our correlation results between the X-ray and optical emission for NGC\,4051, as described in \S\ref{sec:cc_analysis}, remain robust.

\section{Cross-correlation analysis} \label{sec:cc_analysis}

We checked for correlated variability and time lags ($\tau$) among the light curves shown in Figure\,\ref{lc_mrk359} and \ref{lc_ngc4051} using the interpolated cross-correlation function (ICCF) method together with flux redistribution and random subset sampling (FRRSS), as employed by \citealt{Peterson_1998}. We used \texttt{pyCCF} \citep{Sun_2018}, which computes the cross-correlation function for unevenly sampled light curves using linear interpolation and measures the centroid and peak lag of the correlation function. For all light curve pairs, we computed the Pearson correlation coefficient $R(\tau)$ as a function of time lag $\tau$ over a lag grid with spacing $0.1$ and $0.05~{\rm days}$ for Mrk\,359 and NGC\,4051, respectively. For both sources, we tested a lag range of $[-3,3]$ days. We define the centroid lag, $\tau_{\rm cent}$, as the mean weighted lag for values of $R(\tau) > 0.8 R_{\mathrm{max}}$. The time lag uncertainties were estimated using the flux redistribution and random subset sampling (FRRSS) method. FRRSS is a Monte Carlo bootstrapping method which accounts for both sampling and measurement uncertainties by 
resampling the light curves and randomly perturbing the fluxes within their measured uncertainties. The cross correlation function is calculated for each light curve realization and the resulting distribution of centroid time lags is used to estimate the uncertainty on centroid time lag $\tau_{\rm cent}$. Using this method, we performed cross-correlation analysis on the X-ray (NICER) and optical (TESS) light-curve pairs, and among the X-ray light-curve pairs. We also place confidence limits on the observed CCFs using Monte Carlo simulations of light curves assuming no correlation (see Appendix\,\ref{app:simulations}).

\subsection{X-ray/Optical Correlations}

\setlength{\tabcolsep}{6pt}
\renewcommand{\arraystretch}{1.15}
\begin{deluxetable*}{cccccccccccc}[]
\tablecaption{X-ray/optical cross correlation results for Mrk\,359 and NGC\,4051 and comparison with results of other AGN\label{tab:correlation_lags}}
\tablecolumns{12}
\tablewidth{0pt}
\tablehead{
\multicolumn{1}{c}{} &
\multicolumn{4}{c}{This work} &
\multicolumn{1}{c}{\hspace{1em}} &
\multicolumn{6}{c}{Previous Studies} \\
\cmidrule(lr{0.4em}){2-5}
\cmidrule(lr{0.4em}){7-12}
\multicolumn{1}{c}{} &
\multicolumn{2}{c}{Mrk\,359} &
\multicolumn{2}{c}{NGC\,4051} &
\multicolumn{1}{c}{\hspace{1em}} &
\multicolumn{2}{c}{NGC\,7469\textsuperscript{1}} &
\multicolumn{2}{c}{MCG 08-11-11\textsuperscript{2}} &
\multicolumn{2}{c}{Fairall\,9\textsuperscript{3}} \\
\cmidrule(lr{0.4em}){2-3}
\cmidrule(lr{0.4em}){4-5}
\cmidrule(lr{0.4em}){7-8}
\cmidrule(lr{0.4em}){9-10}
\cmidrule(lr{0.4em}){11-12}
\colhead{X-ray Model Parameter} &
\colhead{$R_\mathrm{max}$} &
\colhead{$\tau_\mathrm{cent}\mathrm{~[days]}$} &
\colhead{$R_\mathrm{max}$} &
\colhead{$\tau_\mathrm{cent}\mathrm{~[days]}$} &
\colhead{} &
\colhead{$R_\mathrm{max}$} &
\colhead{$\tau_\mathrm{cent}\mathrm{~[days]}$} &
\colhead{$R_\mathrm{max}$} &
\colhead{$\tau_\mathrm{cent}\mathrm{~[days]}$} &
\colhead{$R_\mathrm{max}$} &
\colhead{$\tau_\mathrm{cent}\mathrm{~[days]}$}
}
\startdata
Total X-ray source Flux &
0.6 &
$-0.6^{+0.2}_{-0.2}$ &
0.5 &
$0.3^{+0.5}_{-0.6}$ &
\hspace{1.5em} &
0.8 &
$-0.8^{+0.5}_{-0.4}$ &
0.85 &
$-1.8^{+0.2}_{-0.2}$ &
0.6 &
$-0.6^{+1.4}_{-0.6}$ \\
Power Law (Hot Corona) Flux &
0.6 &
$-0.6^{+0.2}_{-0.2}$ &
0.5 &
$0.5^{+0.5}_{-0.5}$ &
\hspace{1.5em} &
0.8 &
$-0.7^{+0.5}_{-0.5}$ &
\nodata &
\nodata &
0.5 &
$-0.8^{+2.8}_{-2.0}$ \\
Power Law Photon Index $\Gamma$ &
0.5 &
$-2.1^{+2.4}_{-0.3}$ &
0.6 &
$-0.9^{+0.9}_{-1.0}$ &
\hspace{1.5em} &
0.6 &
$-1.7^{+0.9}_{-1.1}$ &
\nodata &
\nodata &
\nodata &
\nodata \\
\begin{tabular}{@{}c@{}}
Soft Excess Flux \\
\end{tabular} &
0.5 &
$-0.6^{+0.2}_{-0.3}$ &
0.5 &
$0.4^{+0.6}_{-0.6}$ &
\hspace{1.5em} &
0.8 &
$-1.2^{+0.5}_{-0.5}$ &
\nodata &
\nodata &
0.8 &
$-2.4^{+1.4}_{-1.0}$ \\
\enddata
\tablecomments{
For this work, referenced to the TESS optical light curve, the maximum cross correlation coefficient $R_\mathrm{max}$ and lag estimate $\tau_\mathrm{cent}\mathrm{~(days)}$ of the X-ray model components, as calculated using the ICCF and FRRSS method described in \S\ref{sec:cc_analysis}. Negative lags indicate that the variability in X-rays occurs first and is then followed by optical variations after a time lag $\tau$. However, the low correlation strengths observed for both Mrk\,359 and NGC\,4051 do not support a conclusive result regarding the time lags. Previous study results are from \textsuperscript{1}\citealt{2025ApJ...986...81P}, \textsuperscript{2}\citealt{2026MNRAS.tmp...50K}, and \textsuperscript{3}\citealt{2024ApJ...977...77P}, where the Swift UVW2 band was used as the reference light curve for correlation and time lags.}
\label{tab:cc}
\end{deluxetable*}

\begin{figure*}[ht]
\centering
\includegraphics[width=0.8\textwidth]{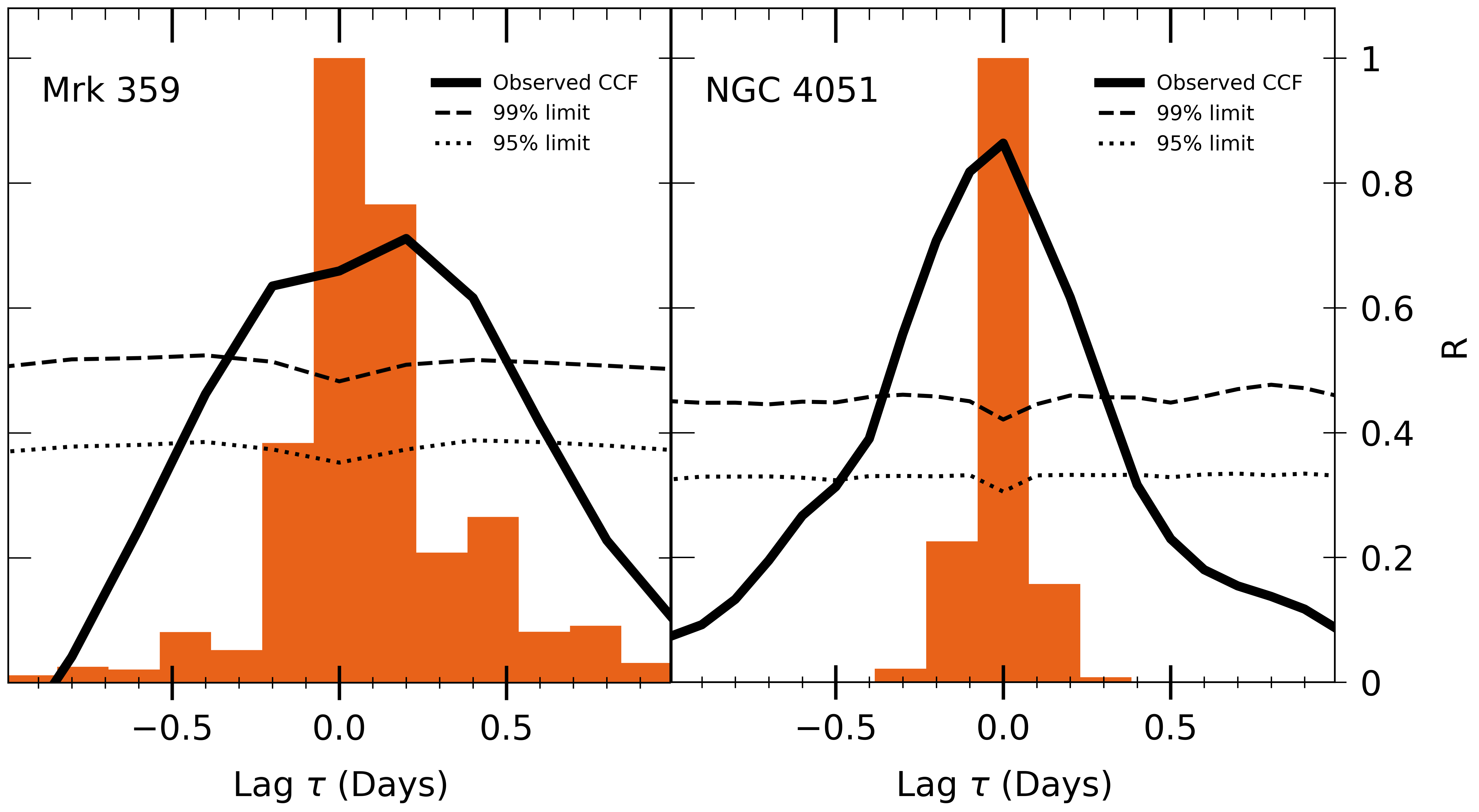}
\caption{\textit{\textbf{Left: }} Centroid distribution for the cross correlation function between the power law flux and the soft excess flux light curves of Mrk\,359. The maximum cross correlation coefficient is $R_{\mathrm{max}}\simeq0.7$ at a lag of $0.0^{+0.3}_{-0.1}$, with respect to the power law flux as the reference light curve. \textit{\textbf{Right: }}Same as left figure but for NGC\,4051. Here $R_{\mathrm{max}}\simeq 0.9$, at a lag of $0.0^{+0.1}_{-0.1}$. The dotted and dashed lines are the 95\,and\,99\,per\,cent confidence limits, respectively, calculated using Monte Carlo simulations of light curves assuming no correlation (see Appendix\,\ref{app:simulations}) 
\label{nth_rel_ccpd}
}
\end{figure*}

For X-ray/optical correlations, we treated the binned TESS light curve as the reference light curve. Since the TESS data have denser, more uniform sampling and smaller uncertainties than the NICER light curves, we interpolate only the TESS data to minimize noise in the resulting cross-correlation function. At each lag $\tau$, the NICER light curve is shifted in time and referenced to the interpolated TESS light curve, and $R(\tau)$ is measured. We do this for the total X-ray flux, the primary X-ray flux, the photon index of the power-law component, and the soft excess flux for each source. For NGC\,4051, since we also have the $u$\,band light curve observed with LCO that provides data in the TESS gap, we also perform the X-ray/optical correlations with the $u$\,band light curve as the reference light curve. 

The resulting X-ray/optical cross-correlation functions and the centroid time-lag distributions for Mrk\,359 and NGC\,4051 are given in the right panels of Figures\,\ref{lc_mrk359} and\,\ref{lc_ngc4051}. The results are also summarized in Table\,\ref{tab:cc}. We found that the correlations between the total X-ray flux and the TESS light curves are low, with maximum cross-correlation coefficients of $R_{\mathrm{max}}$\,$\sim$0.6\,and\,$0.5$ for Mrk\,359 and NGC\,4051, respectively. Cross-correlations with the fluxes of the individual X-ray model components were also low. We obtained $R_{\mathrm{max}}$\,$\sim$0.6\,and\,$0.5$ for the primary power-law flux with respect to the TESS data for Mrk\,359 and NGC\,4051, respectively. Similarly, the correlation between the soft excess and TESS data was low, with $R_{\mathrm{max}}$\,$\sim$\,$0.5$ for both sources. For Mrk\,359, the first half of the light curves ($<59865$) are better correlated with $R_{\mathrm{max}}$\,$\sim$$0.8$ for the total, powerlaw, and the soft excess with respect to the TESS light curve; for the second half, however, the correlation drops to $R_{\mathrm{max}}$\,$<$\,$0.6$. The low correlation coefficients observed for both Mrk 359 and NGC 4051 do not support a conclusive result regarding the time lags. The non-zero $-0.6$ days lag for Mrk\,359 seems to be determined by two high points on the X-ray light curves (MJD 59854.5 and 59868.5) that precede peaks in the TESS light curve (MJD\,59855.2 and 59869.5). 

For NGC\,4051, we find consistent X-ray/optical correlation results with the $u$ band light curve as the reference. The TESS and $u$\,band light curves are also strongly correlated with $R_{\mathrm{max}}$\,$\sim$\,$0.9$ and the TESS data varying at a lag of $0.9 \pm 0.2\rm\,days$ with respect to the $u$\,band variations. We,\,therefore,\,report cross-correlation results in Table\,\ref{tab:cc} with respect to TESS data only. For NGC\,4051, we find consistent correlation coefficients with \textit{XMM-Newton} observations of \citealt{2013MNRAS.429...75A}, where they found a UV/X-ray correlation coefficient of $R$$\sim$0.5, with the UV lagging the X-rays by $\sim$3\,ks.

\subsection{Primary X-ray/Soft excess correlation}

We also checked for correlated variability between the primary X-ray flux and the soft excess. We chose the primary X-ray light curve as the reference light curve. Here, since both light curves have the same sampling and comparable uncertainties, we interpolate each light curve in turn, calculate the cross-correlation functions, and then obtain the final CCF by averaging the two results.

We give the primary X-ray/soft-excess cross correlation results for both sources in Figure\,\ref{nth_rel_ccpd}. For Mrk\,359, we found $R_{\rm{max}}$\,$\sim$0.7 signifying a higher correlated variability between the X-ray model components compared to correlations obtained with the optical variations. For NGC\,4051, we found $R_{\rm{max}}$\,$\sim$0.9 signifying a strong correlated variability between the primary X-rays and soft excess flux. For both sources, the lags obtained are consistent with $0$ days.

We also examined the correlation between the power law photon index $\Gamma$ and the soft excess flux. For Mrk\,359 there is no correlation with $R_{\rm max} \sim 0.3$. For NGC\,4051, the soft excess flux increases as the primary continuum steepens, with a correlation of $R_{\max} \sim 0.7$, similar to the correlation with the primary X-ray flux.

\section{Discussion} \label{Discussion}

Our timing analysis of Mrk\,359 and NGC\,4051 shows that the optical variations are only weakly correlated with the total X-ray flux, the primary X-ray flux, and the soft-excess flux. In contrast, the soft excess flux is more strongly correlated with the primary X-ray flux in both sources. Any physical interpretation must explain both the presence of a soft excess component and why the soft excess weakly correlates with the optical emission and more coherently with the primary X-ray continuum.

Our discussion is organized around these two observational constraints. First, we discuss possible reasons for the weak correlation between the X-ray and optical light curves. Second, we investigate how the multiwavelength variability of AGN inform our understanding of the soft excess. Finally, we discuss mechanisms other than X-ray reprocessing that can drive the UV/optical variability in Mrk\,359 and NGC\,4051. 

\subsection{Weakly correlated X-ray and optical light curves} \label{x-ray_optical_discussion}

In the standard linear reprocessing model for AGN, the X-rays irradiate the accretion disk and produce reprocessed emission in the UV/optical band. Variations in the driving X-ray light curve ($F_{\rm X}$) maps on to the variations in the UV/optical reprocessed light curve ($F_{r}$) after a light-travel time delay $\tau$. This process ---with the use of a transfer function $\Psi(\tau)$---is encoded in the following formalism:
\begin{equation}
    \Delta F_r(t) = \int_{0}^{\tau_{\max}}\Psi(\tau)\,\Delta F_X(t - \tau)\, d\tau
    \label{rmequation}
\end{equation}
where the transfer function $\Psi(\tau)$ captures the response of the disk to the driving flux variability. In this scenario, the reprocessed UV/optical light curves will be a smooth version of the X-ray light curve, and the observed X-ray--UV/optical correlation should be strong.

However, we found that there is no strong correlation between total X-ray flux and optical counts variability, with $R_{\mathrm{max}} =$\,0.6 and 0.5 for Mrk\,359 and NGC\,4051, respectively. The primary X-ray flux modeled using a power-law component also do not show strong correlated variability with the optical counts, with $R_{\mathrm{max}} =$\,0.6 and 0.5 for Mrk\,359 and NGC\,4051, respectively. Further, according to Equation\,\ref{rmequation}, if the driving source flux varies like a $\delta$-function flare, the reprocessed emission should also vary similarly. We find no evidence to support Equation\,\ref{rmequation} in our case. For example, the total X-ray flux for NGC\,4051 increased from $0.7\times10^{-10}$ to $1.2\times10^{-10}$ and dropped back to a similar flux in $\sim$0.5$\mathrm{\,day}$, at MJD $59659.5$. However, there is no clear response in the optical light curve; instead, the optical counts show a coherent variation from MJD 59656 to 59662. Thus, the response of fast X-ray impulses are not present in the observed optical light curves. Together, these results suggest that a static geometry central X-ray corona illuminating the accretion disk is not the mechanism driving the optical variability on timescales of days in Mrk\,359 and NGC\,4051.

Several scenarios may explain the X-ray/optical correlation results. First, a low X-ray/UV correlation can arise even in the case of X-ray reprocessing when the dynamic variability of the central X-ray corona is considered. \citealt{2022ApJ...941...57P} showed that random variations in the height of a lamppost corona between $5$ and $30\,R_{\mathrm{g}}$ every 15 days of simulated light curves produce a range of X-ray/UV correlation coefficients, with a preference for low values of $R_{\mathrm{XW2}}\sim0.45$--$0.8$. Their dynamic corona model also preserved the strong UV/optical correlations expected from X-ray reprocessing and observed in many AGN. Therefore, short timescale changes in the coronal geometry can naturally reduce the observed X-ray--UV/optical correlation, even if fast X-ray reprocessing drives the UV/optical variability. However, if the soft excess is dominated by reflected emission, even in a dynamic corona model we expect its variability to remain correlated with the optical variability, which is not observed.

A second possibility suggested by \citealt{2024ApJ...965L..29S,2025ApJ...984...19S} is that intrinsic disk fluctuations from the MRI or convection dominate the UV variability when the X-ray flux is small or when X-ray absorption opacity is small.  However, X-ray reflection models include the combined effects of scattering  and absorption in the disk, and so we would still expect to find a correlation if the soft excess is produced by X-ray reflection. We must therefore conclude that either X-rays are unimportant for driving the UV/optical variations in these AGN, or that their soft excesses are not produced by X-ray reflection.

A final explanation is that the weak X-ray/optical correlations could be explained by an intervening vertical structure between the hot corona (responsible for primary X-rays) and the UV/optically emitting part of the accretion disk. Such a structure likely needs a large vertical scale height, which could be attributed to atomic processes such as winds driven by UV lines \citep{2014MNRAS.438.3024L} or a vertically extended ``warm corona" \citep{2012MNRAS.420.1848D,2017MNRAS.470.3591G}.
Either way, the intervening structure shields the accretion disk from X-rays\,$>$\,2.0\,keV originating close to the central SMBH and dilutes the correlation between fast X-ray and UV/optical variability \citep{2017MNRAS.470.3591G}. 

\subsection{Relationship between UV/optical variability and the soft excess}

We investigate how the multiwavelength variability of AGN inform our understanding of the soft excess.

In a simple linear X-ray reprocessing model, a fraction of the driving primary X-ray flux is absorbed by the disk while the remainder of the flux is reflected. The reflected spectrum can contribute to the soft excess flux, either through blurring of fluorescent lines below $\sim$\,$2$$\mathrm{\,keV}$ or free-free emission in high density disks ($\gtrsim 10^{18}\,\mathrm{cm^{-3}}$; \citealt{2016MNRAS.462..751G}). If the soft excess is dominated by X-ray reflection, the soft-excess and optical light curves are expected to vary coherently, since both respond to the same driving primary X-ray light curve according to Equation\,\ref{rmequation}. For NGC\,4051, the soft excess can be modeled with \texttt{RELXILLCP}; for Mrk\,359, the reflection parameters could not be constrained with \texttt{RELXILLCP}, so we instead used a phenomenological blackbody to measure the soft-excess flux. The timing results therefore provide an additional test of whether the soft excess in either source is consistent with a reflection-dominated origin. To assess whether the soft excess in Mrk\,359 and NGC\,4051 is consistent with an X-ray reflection origin, we first compare our results with observational constraints from NGC\,7469, MCG-08-11-11, and Fairall\,9.

A time-resolved spectral modeling campaign with NICER on NGC\,7469 showed that the primary X-ray flux variability correlates strongly ($R$\,$\simeq$\,0.8) with the UV variability \citep{2025ApJ...986...81P}. They also measured a strong correlation ($R$\,$\simeq$\,0.9) between the primary X-ray and soft excess fluxes, along with a strong correlation ($R$\,$\simeq$\,0.8) between the soft excess and UV variability. The coherent variability among the three components---the primary X-rays, the soft excess, and the UV flux---suggests that a standard X-ray reprocessing scenario can explain both the soft excess and the spectral-timing properties of NGC\,7469. MCG+08-11-11\,shows one of the strongest correlations between the 0.3--10\,keV X-ray and UV flux  ($R$\,$\simeq$\,0.85; \citealt{2026MNRAS.tmp...50K}). While the strong correlation supports the X-ray reprocessing framework, MCG+08-11-11 does not have a detectable soft excess. The lack of a soft excess might then be explained by a weak reflection component, possibly from a relatively low disc density or a large inner radius caused by disk truncation or low black hole spin. 

Fairall\,9 is a more complicated case, where the soft-excess and UV fluxes are strongly correlated ($R\simeq 0.8$), but the primary X-rays show only weak correlations with both the soft excess and the UV emission. These correlations could be consistent with X-ray reflection if the observed primary X-ray light curve is not the driving light curve for variations at lower energies. Why this would be the case is not clear, though it could be due to anisotropic emission, an intervening structure between the primary X-rays with partial obscuration of the  disk, or multiple sources of variability for the X-rays operating on different timescales. \cite{2024ApJ...977...77P} instead interpret the soft excess in Fairall\,9 as emission from a warm corona. In this scenario, the warm corona produces the soft X-rays and extreme-UV emission that can drive the longer-wavelength UV/optical variability, while shielding the disk from the hard X-ray flux produced by the hot corona. \citealt{2024MNRAS.530.4850H} suggested that the soft excess in Fairall\,9 is produced by a warm corona, and that intrinsic disk fluctuations drive the correlated optical/UV and soft excess variability. The UV and X-ray emission are only weakly correlated due to a combination of varying disc seed photons entering the corona and X-ray photons reprocessing off the disc.

Unlike the three sources discussed above, the X-ray components in Mrk\,359 and NGC\,4051 show no strong correlations with the optical flux, while the primary X-rays and the soft excess emission show stronger correlated variability. If the soft excess is dominated by reflected emission, a simple X-ray reflection/reprocessing model predicts that the soft excess and the reprocessed optical flux should vary coherently. Thus, the lack of correlated optical variability in either source does not directly provide direct evidence supporting a reflection origin of the soft excess.

We cannot completely rule out a reflection origin for the soft excess, as it is possible that the reflection emission originates from the inner regions of the disk, while the outer disk is shielded from the X-rays produced in the inner regions by an intervening vertical structure. The stronger correlated variability between primary X-rays from the hot corona and the soft excess would then be a result of variations driven in reflected X-ray spectrum by a varying incident X-ray spectrum. As discussed in \S\,\ref{x-ray_optical_discussion}, such an intervening component can dilute the correlation between the X-rays and the optical flux variations. The X-ray variability of the two sources could therefore be similar to that observed in NGC\,7213, except that the optically emitting disk is not directly illuminated by the hot corona. With a cadence of 0.5 days, we cannot test this hypothesis, because the lag between the primary X-rays and the reflected component would be much shorter (of order $10^{2}$--$10^{3}\,\rm s$ at $10$--$100\,R_{\rm g}$ in NGC\,4051). However, such an intervening structure will have a distinct spectral component of its own, which may be problematic for pure-reflection models of the soft excesses.

\subsection{Other viable mechanisms for optical variability} \label{origin_optical_variability}

We discuss mechanisms other than X-ray reprocessing that can drive the UV/optical variability in Mrk\,359 and NGC\,4051 on timescales of days. One possibility is that the driving light curve is in the extreme-UV, originating near the inner edge of the disk, which may also be the outer extent of a Comptonizing warm corona \citep{2017MNRAS.470.3591G, 2023MNRAS.521..251H,2024MNRAS.530.4850H}. A second possibility is that the 600--1000\,nm TESS band, spanning wavelengths equivalent to the SDSS \textit{r}, \textit{i}, and \textit{z} filters, contains reprocessed emission from multiple components. The multiple components could include disk, winds, and diffuse continuum emission from the BLR, each with different levels of smoothing. These components with different time lags and timescales of variability would dilute any fast variations driven by X-ray irradiation \citep{2001ApJ...553..695K, 2018MNRAS.481..533L,2023MNRAS.519.2754S}. However, as seen for NGC\,4051, the TESS and $u$\,band light curves are strongly correlated, which implies that the TESS band is tracing the same disk variability as the $u$\,band.

It is also possible that the optical variability is driven by mechanisms other than reprocessing. One such proposed mechanism is intrinsic disk variability produced by mass accretion rate fluctuations, similar to the process suggested to explain the X-ray variability in stellar-mass black hole binaries \citep{1997MNRAS.292..679L, 2006MNRAS.367..801A, 2011MNRAS.415.2323I, 2024MNRAS.530.4850H, 2025MNRAS.536.3284U}. In this model, the local accretion rate at each disk annulus fluctuates, and fluctuations generated at larger radii propagate inward, adding progressively more variability in the inner regions of the disk. For a standard thin disk, the fastest timescale on which accretion rate can change substantially is generally associated with the viscous timescale \citep{1973A&A....24..337S}. This propagating fluctuations model has been suggested for stellar-mass black hole binaries, where the physical timescales are short, and it matches the variability in the high/soft state \citep{2006MNRAS.367..801A}. In AGN disks, however, the viscous timescale at the optical-emitting radii is by many orders of magnitude longer than the observed UV/optical variability timescales, making the local accretion rate fluctuations difficult to explain the day-to-week timescale UV/optical variability in AGN. Nevertheless, phenomenological models based on propagating fluctuations through disk annuli can explain key observational results, including the X-ray-to-UV/optical disconnect \citep{2024MNRAS.530.4850H}.

A second possibility is that the observed UV/optical variability is from a strongly inhomogeneous disk \citep{2011ApJ...727L..24D}. In this model, the intrinsic disk variability is a combined effect of many independent, stochastically fluctuating regions rather than coherent variations in the disk, such as due to propagating fluctuations. \citealt{2011ApJ...727L..24D} showed that large localized temperature fluctuations, with temperatures at a radius varying by a factor of 3-5, can explain the stochastic properties of quasar variability and solve the discrepancy in accretion disk sizes. However, the physical mechanism and the characteristic timescale for such stochastic temperature fluctuations are not understood. Fluctuations due to magnetic confinement and reconnection processes may drive these local variations \citep{2009ApJ...691...16H, 2013ApJ...778...65J}. 

If the UV/optical variability is not caused by X-ray reprocessing, it suggests that the observed lags at these wavelengths of other AGN may not be due to light travel times. If this is the case, previous reports of AGN accretion disk temperature profiles may be called into question, since the physical size of the disk is not represented by the time delay between bands.

The observed variability will become more complex if X-ray reprocessing contributes in addition to any such intrinsic disk variability. In this scenario, the UV/optical light curves could contain both a slow component dominated by intrinsic variability and a fast component dominated by reprocessing. The X-ray light curve may exhibit slow variations dominated by intrinsic variability, and fast variations due to coronal fluctuations and reprocessing. Future tests with frequency-resolved time lags will be required to separate any such fast and slow variability processes and to study the coherence between X-ray and UV/optical light curves.

\section{Summary and Conclusions} \label{sec: conclusions}

We observed two AGN, Mrk\,359 and NGC\,4051, in the X-ray and optical bands using NICER and TESS over a period of $\sim$$30\,\rm days$. The X-ray spectra of both sources reveal a soft excess above the extrapolation of the hard X-ray power-law continuum. We performed time-resolved spectroscopic fits to the NICER data to obtain separate broadband fluxes for the primary X-ray and soft excess components. The optical light curves were obtained from TESS data using difference imaging photometry. Our main results are summarized as follows.

\begin{enumerate}

    \item For Mrk\,359, we used a phenomenological \texttt{blackbody} model to fit the soft excess, together with a \texttt{powerlaw} component for the hard X-ray continuum. During our observations, Mrk\,359 was in a state where the ratio of soft excess to primary X-ray continuum flux was smaller by a factor of $\sim$2 compared to that observed by \citealt{2020A&A...640A..99M}. The cross-correlation analysis shows no strong correlations between the optical light curve and the total X-ray, primary X-ray, and soft-excess fluxes. 

    \item For NGC\,4051, the observations reveal a prominent soft excess, which we fit using the high-density accretion disk X-ray reflection model \texttt{RELXILLCP}. The cross-correlation analysis again shows low correlations between the optical light curve and the X-ray spectral components. 

    \item The weak correlation between the observed X-ray emission and the TESS optical light curves does not favor X-ray reprocessing for the origin of UV/optical variability. These results may also pose challenges to reflection models of the soft excesses.

    \item Possible explanations of these results are that the optical variability may be driven by non-reprocessing mechanisms, such as propagating accretion rate fluctuations or inhomogeneous disk variability associated with magnetic processes. Alternatively, the driving light curve may originate in the extreme UV. In this case, an explanation is needed for why the optically emitting disk does not receive X-ray flux. One possibility might be a vertically extended structure that shields  the disk from the hard X-rays. It is tempting to associate such a structure with a ``warm corona" spectral component, which may share the same source of fast variability as the hard X-rays and could potentially account for the soft excess emission.
    
\end{enumerate}

\begin{acknowledgments}

This material is based upon work supported by the National Aeronautics and Space Administration under Grant  No. 80NSSC24K0168 and  Grant No. 80NSSC24K0187. This paper includes data collected with the TESS mission, obtained from the MAST data archive at the Space Telescope Science Institute (STScI). Funding for US Institutions for the TESS mission is provided by the NASA Explorer Program. STScI is operated by the Association of Universities for Research in Astronomy, Inc., under NASA contract NASA 5–26555.

\end{acknowledgments}

\section*{Data Availability}
The TESS full-frame images (FFIs) used in this work are publicly available from the Mikulski Archive for Space Telescopes (MAST), Sector\,57 for Mrk\,359 at \dataset[doi:10.17909/zamj-tf92]{https://doi.org/10.17909/zamj-tf92}, and Sector\,49 for NGC\,4051 at \dataset[doi:10.17909/mqp3-d907]{https://doi.org/10.17909/mqp3-d907}. The NICER observations used in this work are publicly available through NASA's High Energy Astrophysics Science Archive Research Center (HEASARC): \href{https://heasarc.gsfc.nasa.gov}{https://heasarc.gsfc.nasa.gov}. The NGC\,4051\,\textit{u}\,band light curve data obtained with the Las Cumbres Observatory (LCO) global robotic telescope network and presented in \citealt{2026MNRAS.548ag642M} are publicly available on Zenodo \citep{doi:10.5281/zenodo.17242993}.

\software{HEASoft \citep{Blackburn_1995, 2014ascl.soft08004N}, XSPEC \citep{Arna96}, emcee \citep{emcee},  corner \citep{Foreman-Mackey2016}, PyCCF \citep{Peterson_1998,Sun_2018}}

\restartappendixnumbering
\appendix

\section{Simulating Light Curves} \label{app:simulations}

We place confidence intervals on the observed cross correlation functions (CCFs) between the X-ray and optical light curves, and between the X-ray light curves, using Monte Carlo simulations of uncorrelated light curve pairs. We used the method of \citealt{1995A&A...300..707T} to generate light curves using an appropriate power spectral density (PSD), which were then sampled at observed timestamps, using the corresponding exposure windows. 

For NGC\,4051, the simulated X-ray light curves were generated from a bending power-law PSD with a low-frequency slope of $-1$, a high frequency slope of $-2$, and a break frequency of $2\times10^{-4}\rm\,Hz$ \citep{2011MNRAS.413.2489V}. For Mrk\,359, we assumed the same PSD as NGC\,4051 because no previous PSD measurement has been reported and the two sources have similar black hole masses. The TESS and $u$-band PSDs were modelled using an unbroken power law with slope of $-2$. 

We generated $10^{4}$ pairs of uncorrelated X-ray and TESS/$u$\,band light curves and computed their CCFs using the ICCF method described in \S\,\ref{sec:cc_analysis}. The distributions of simulated CCFs were used to determine the 90,\,95,\,and\,99\,per cent confidence limits. The\,99\,per\,cent limit is R$\lesssim$\,0.6 for X-ray/optical correlations. For the cross-correlation between the primary X-ray continuum and the soft excess light curves, we followed the same procedure using pairs of uncorrelated simulated X-ray light curves.

\section{Supplementary Spectral Fitting Results for NGC\,4051} \label{app:ngc4051}

\begin{figure*}[]
\centering
\includegraphics[width=\textwidth]{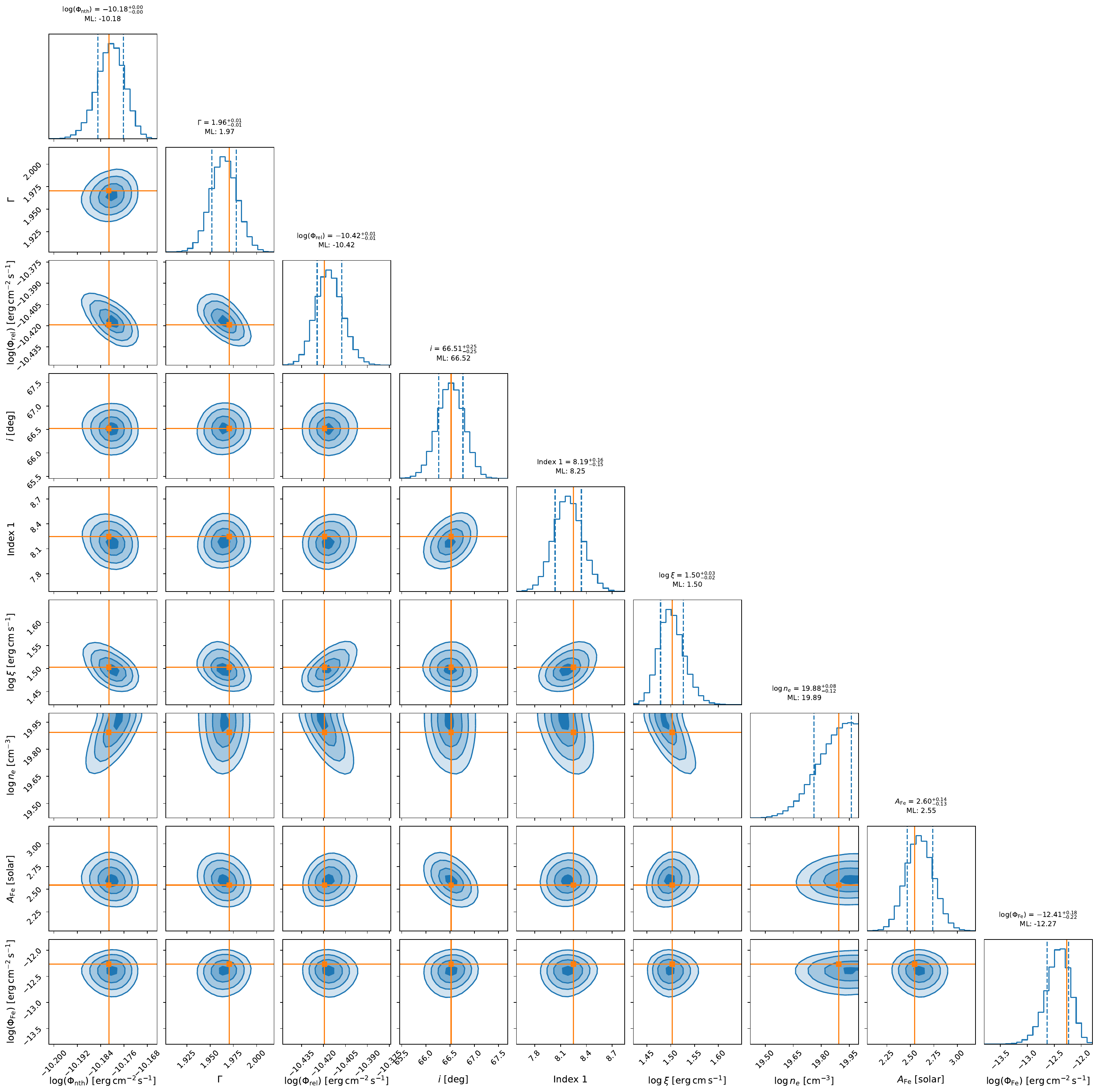}
\caption{Corner Plot for NGC\,4051 observation taken on MJD $59653.89$, whose spectrum is shown in Figure\,\ref{single_epochs_spectrum_ngc4051}. The on-diagonal histograms represent the 1-d posterior for parameters, with the vertical orange line showing the maximum-likelihood parameter along with 68.3\% confidence intervals. The title of each on-diagonal histogram presents the median value along with confidence intervals and the corresponding maximum-likelihood parameter. Different blue contours in the off-diagonal elements represent the $0.5\sigma, 1\sigma, 1.5\sigma, \mathrm{and}\ 2\sigma$ confidence intervals for 2-d Gaussian distribution pairs. 
\label{corner_NGC4051}
}
\end{figure*}

In Figure\,\ref{corner_NGC4051}, we present an example posterior distribution from our fitting procedure in \S3 for a NGC\,4051. The spectral model, in \texttt{XSPEC} terminology, used to obtain the posterior is \texttt{tbabs\,$\times$\,(cflux\,$\times$\,nthcomp\,+\,cflux\,$\times$\,relxillCp\,+\,cflux\,$\times$\,zgauss)}. The observation was taken on MJD $59653.89$ and its spectrum is shown in Figure\,\ref{single_epochs_spectrum_ngc4051}.

\begin{figure*}[]
\plotone{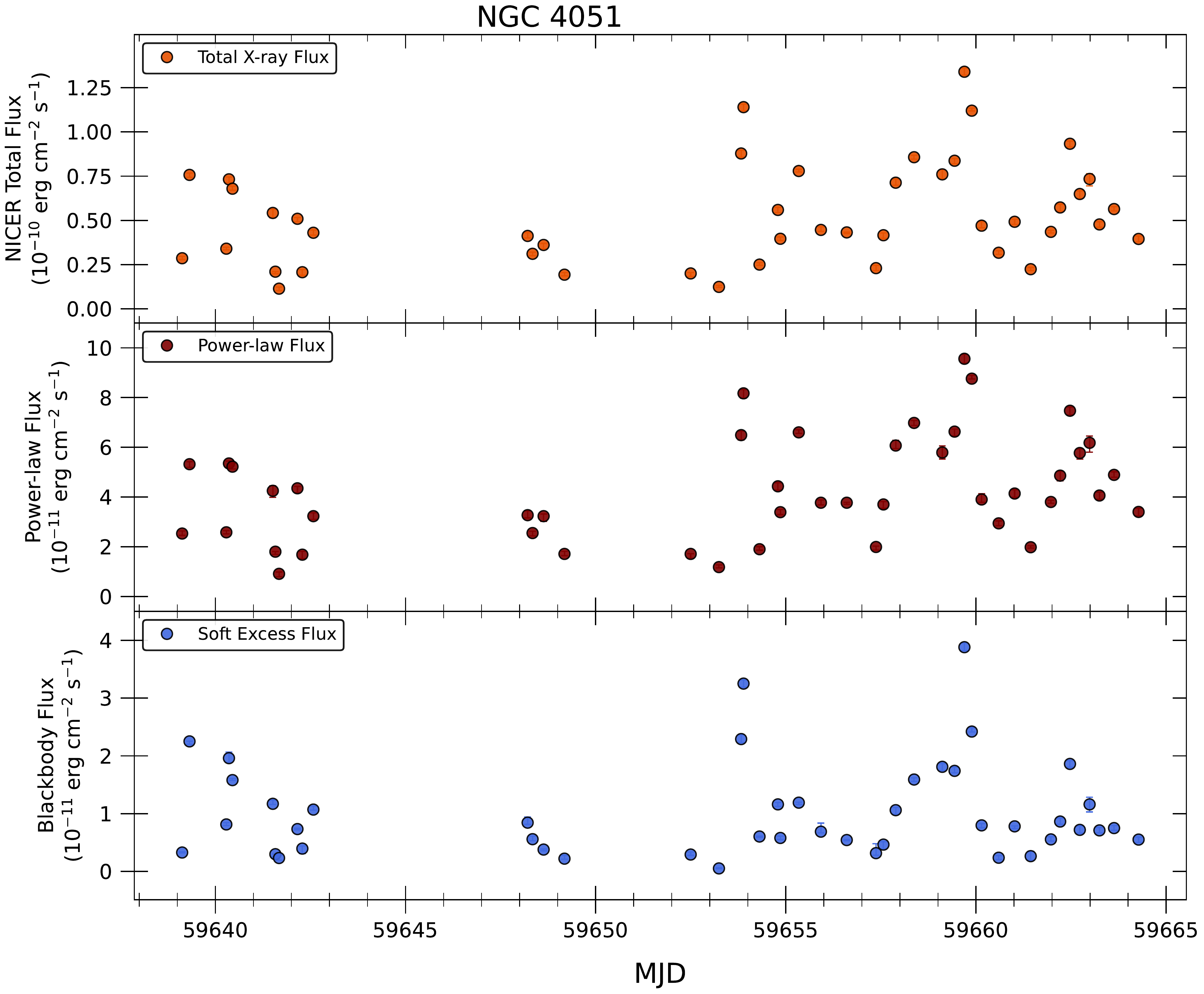}
\caption{Same as Figure\,\ref{lc_ngc4051}, but here we plot the unabsorbed source fluxes obtained using a spectral model assumed to be a \texttt{powerlaw} with photon index $\Gamma=2.0$ (fixed) and \texttt{blackbody} with $kT = 0.12\,\rm\,keV$ (fixed), both modified by two variable absorbers modelled using \texttt{zxipcf}. The error bars were calculated using the \texttt{XSPEC error} command and represent the $68\%$ confidence interval. The shape of the light curve obtained using the absorbed \texttt{powerlaw} spectral model is the same as the one presented in Figure\,\ref{lc_ngc4051}, where the photon index was left free in the range 1.0--3.0. This result implies that the strong X-ray flux variability of NGC\,4051 is dominated by intrinsic variations in the source flux rather than due to any variable absorption that modifies the power-law shape. Thus, our correlation results between the X-ray and optical emission for NGC\,4051, as described in \S\ref{sec:cc_analysis}, remain robust. 
\label{lc2_ngc4051}}
\end{figure*}

In Figure\,\ref{lc2_ngc4051}, we present the X-ray light curves of NGC\,4051 obtained using a spectral model assumed to be a \texttt{powerlaw} with photon index $\Gamma=2.0$ (fixed) and a \texttt{blackbody} with $kT = 0.12\,\rm\,keV$ (fixed), both modified by two variable absorbers modelled using \texttt{zxipcf}. The spectral model, in \texttt{XSPEC} terminology, used to obtain the light curves is \texttt{tbabs$\times$\,zxipcf$ \times$\,zxipcf$\times$\,(cflux$\times$\,powerlaw\,+\,cflux$\times$\,bb)} and the procedure is described in \S\,\ref{subsec:NGC4051}.

\bibliography{sample702}{}
\bibliographystyle{aasjournalv7.1}



\end{document}